\documentclass[a4paper,12pt]{article}
\pdfoutput=1 
\usepackage{graphicx}
\usepackage{bm}
\def\be{\begin{equation}}
\def\ee{\end{equation}}
\def\bea{\begin{eqnarray}}
\def\eea{\end{eqnarray}}
\usepackage{latexsym}
\usepackage{dcolumn}
\usepackage{amsmath}
\usepackage{epsfig,amssymb,euscript,mathrsfs}
\usepackage{array,calc,epsfig}

\usepackage[sans]{dsfont} 

\usepackage[scaled=1.2]{dsserif} 

\usepackage{authblk}

\usepackage[noadjust]{cite}

\usepackage{chngcntr}

\usepackage[pdfpagelabels]{hyperref}

\usepackage{color}

\newcommand{\Tr}{{\rm Tr}}

\renewcommand{\to}{\rightarrow}

\def\be{\begin{equation}}
\def\ee{\end{equation}}
\def\ba{\begin{eqnarray}}
\def\ea{\end{eqnarray}}
\def\nb{\nonumber}

\def\b{\beta}

\counterwithin*{equation}{section}

\def\({\left (}
\def\){\right )}

\let\benn\[
\let\eenn\]

\def\[{\left [}
\def\]{\right ]}

\let\oldlgraf\{ 
\renewcommand{\{}{\left \oldlgraf}

\let\oldrgraf\}
\renewcommand{\}}{\right \oldrgraf}

\newcommand{\rmd}{\,\mathrm{d}}

\usepackage{xfrac} 
\usepackage{float} 
\usepackage{subfig}

\newlength\dlf  

\usepackage{booktabs}
\usepackage{comment}
\usepackage{cases} 
\usepackage{empheq}
\usepackage{cancel}

\title{\bfseries\LARGE On the asymptotic density of states\protect\\  in solvable models of strings}

\author[a,b]{Tommaso Canneti
}

\affil[a]{Istituto Nazionale di Fisica Nucleare, Sezione di Firenze %
\protect\\ Via G. Sansone 1; 50019 Sesto Fiorentino (Firenze), Italy}
\affil[b]{Dipartimento di Fisica e Astronomia, Universit\`a di Firenze %
\protect\\ Via G. Sansone 1; 50019 Sesto Fiorentino (Firenze), Italy}

\date{\small canneti@fi.infn.it}

\begin{document}
\baselineskip=15.5pt
\pagestyle{plain}
\setcounter{page}{1}
\newfont{\namefont}{cmr10}
\newfont{\addfont}{cmti7 scaled 1440}
\newfont{\boldmathfont}{cmbx10}
\newfont{\headfontb}{cmbx10 scaled 1728}
\renewcommand{\theequation}{{\rm\thesection.\arabic{equation}}}
\renewcommand{\thefootnote}{\arabic{footnote}}

\maketitle\thispagestyle{empty}
\begin{abstract}
We present a closed formula for the asymptotic density of states for a class of solvable superstring models on curved backgrounds. 
The result accounts for the effects of the curvature of the target space in a concise way. 
\end{abstract}

\newpage

\tableofcontents

\section{Introduction}


String theory studies how elementary particles and fundamental interactions emerge from the assumption that the building blocks of what surrounds us are actually one-dimensional strings of length scale $\sqrt{\alpha'}$, rather than point-like particles.
As a consequence, physical states arise as stringy vibrational modes and their degeneracy is well-known to grow exponentially with the energy, if the latter is large enough. Notoriously, this leads to the existence of a critical value for the temperature above which the string partition function diverges, a phenomenon known also in gauge theories \cite{Hagedorn:1965st, Frautschi:1971ij, Carlitz:1972uf}. We are talking about the so-called \emph{Hagedorn temperature} $T_H$.

The computation of the asymptotic density of stringy states is a well-understood problem in flat space \cite{Huang:1970iq, Sundborg:1984uk, Bowick:1985az, Tye:1985jv, Matsuo:1986es}. In principle, it can be extended to superstrings in curved space (e.g., see~\cite{Russo:2002rq}). Nevertheless, the quantization of the string in such a background is required. Besides flat space, it can be performed just on some particular supergravity solutions. For instance, this applies to the Penrose Limit of $AdS$-geometries known as \emph{pp-wave backgrounds}~\cite{Tseytlin:1995fh, Metsaev:2001bj, Metsaev:2002re, Blau:2002mw, Horowitz:1989bv, Horowitz:1990sr,Ehlers:1962zz, Stephani:2003tm, Bicak:2000ea, Amati:1988ww, Amati:1988sa,Penrose1976, Kowalski-Glikman:1984qtj, Gueven:2000ru,Blau:2002dy, Figueroa-OFarrill:2001hal, Blau:2001ne, Berenstein:2002jq}. 
Generically, we can refer to a supersymmetric world-sheet theory featuring the same number of bosonic and fermionic modes, which are massive as a consequence of the curvature of the target space. 

In this work, we aim to compute the density of states describing the degeneracy of highly excited states in the spectrum of solvable closed superstring models on curved backgrounds without running dilaton. By solvable, we mean that the string can be quantized exactly using canonical quantization in a light-cone gauge. In other words, the world-sheet sigma model is solvable in terms of free oscillators whose dynamics is assumed to be described by a light-cone Hamiltonian. Here, the latter is supposed to be quadratic. 

In particular, our goal is to derive a closed formula which takes into account the corrections with respect to the flat space result.\footnote{In other words, we are addressing one of the left open problems of \cite{Russo:2002rq}.}
What we expect to get for the density of states $d$, as a function of the eigenvalue $\mathcal E$ of the oscillatory part of the world-sheet Hamiltonian in some units, is
\be \label{dansatz}
d \approx d_0 \,\, e^{A\,\sqrt{\mathcal E}}\, \mathcal E^{-B} \[1+\mathcal O \(1/\sqrt{\mathcal E}\)\] \, , \quad \mathcal E \to +\infty \, .
\ee
Here, $d_0$, $A$ and $B$ are constant parameters which depend on the model under examination. We expect them to receive corrections which depend on the mass and the number of the world-sheet modes.

The value of $B$ is important for establish the thermodynamic properties of a gas of strings in the near Hagedorn regime. 
Strings thermodynamics has been deeply discussed through the years, both in flat and in curved space (see, e.g., \cite{Mitchell:1987th, Bowick:1985az, Sundborg:1984uk, Tye:1985jv, Polchinski:1985zf, OBrien:1987kzw, McClain:1986id, Alvarez:1985fw, Alvarez:1986sj, Atick:1988si, Kogan:1987jd, Sathiapalan:1986db, Deo:1989bv, Deo:1988jj, Bowick:1989us, Brandenberger:1988aj, Mertens:2015ola, PandoZayas:2002hh, Greene:2002cd, Brower:2002zx, Grignani:2003cs, Aharony:2003sx, Barbon:2004dd, Dienes:1998hx, Grignani:1999sp, Mitchell:1987hr, Salomonson:1985eq, Lowe:1994nm, Abel:1999rq, Abel:1999dy, Barbon:1998ix, Barbon:1998cr, Barbon:2001di, Matsuo:1986es, Alvarez:1984ee, Alvarez:1986sj, Deo:1991mp, Deo:1991af, Frey:2023khe, Giddings:1989xe, Jain:1997ga}). For instance, in \cite{Dienes:1998hx}, besides the string partition function, the authors studied the behavior of other relevant thermodynamic quantities as one approaches the Hagedorn temperature from below. More precisely, they focused on the free and the internal energy of a gas of strings, as well as on its specific heat. The conclusion is that they all diverge if $B$ is less than or equal to a particular critical value, while they remain finite if $B$ is greater than it. In the former case, an infinite amount of energy would be necessary to raise the temperature of the system above $T_H$, which gets the meaning of \emph{limiting} temperature for this reason. On the contrary, in the latter case there is no such a barrier and we can think of the system as transiting from one phase to another, $T_H$ being the onset of a phase transition. Let us stress that this analysis has been performed in the canonical ensemble. Nevertheless, the energy fluctuations may be large in the Hagedorn regime. Therefore, a microcanonical treatment would be certainly more accurate. See \cite{Brower:2002zx,Aharony:2003sx} for details. Here, by ``limiting'' we mean that it is not possible to have matter at equilibrium with $T>T_H$ in the thermodynamic limit. This does not mean, of course, that it is not possible to have matter out of equilibrium.

As we will see, the curvature of the background does not affect $A$. Nevertheless, it significantly modifies the value of $B$. Therefore, the question is whether the curvature effects are large enough to modify the phenomenology of the system near the Hagedorn temperature or not. To the best of our knowledge, the value of $B$ in curved space has never been computed from first principles, and the same applies to $d_0$ and to the other $\mathcal O\(1/\sqrt{\mathcal E}\)$-corrections. This is the main motivation for us to address the problem.

After laying the foundations of the computation in section \ref{sec:genform}, we briefly review the flat space case in section \ref{sec:flat}, stressing some inspiring remarks which will be crucial for the generalization to the curved case. The latter has been explored in section \ref{sec:ppwave}, where we consider solvable closed superstring models on a generic background without running dilaton. In section \ref{sec:stringstherm}, we review some aspects of string thermodynamics in curved space applying our findings to a large class of concrete examples. For the reader convenience, we collect our main novel results in section \ref{sec:conclusions}, wrapping out some conclusions. Complementary details can be found in the appendices.

\section{A general formula for any regime}
\label{sec:genform}

Let us consider a closed superstring embedded in a Lorentzian ten-dimensional background. For the time being, let us suppose that none of the directions are compact. 

In general, we can assume that the world-sheet spectrum displays eight bosonic modes of masses $b_i \, \mu$, $i=1,...,8$, and eight fermionic modes\footnote{We are considering a Lorentzian background at zero temperature. Therefore, the world-sheet fermions are taken periodic along the spatial world-sheet direction.} of masses $f_i \, \mu$, $i=1,...,8$. Here, $\mu$ stands for a (dimensionless) mass-scale.  As a prototype example of this class of theories one can think about strings on pp-waves backgrounds.
Let us stress that, in the absence of a running dilaton, the theory is conformal anomaly free if the sum of the squared masses of the bosonic propagating modes is equal to the sum of the squared masses of the fermionic propagating modes (e.g., see \cite{Drukker:2000ep,Gautason:2021vfc}), that is
\be \label{massmatchppwave}
\sum_{i=1}^8 b_i^2 = \sum_{i=1}^{8} f_i^2 \, .
\ee

For solvable (quadratic) models, the oscillatory part of the light-cone Hamiltonian can be written as
\be \label{Hppwave}
\mathcal{H} = \sum_{k \in \mathds Z} \sum_{i=1}^8 \Bigl ( |\omega_{k i}^B| N_{k i}^B + |\omega_{k i}^F| N_{k i}^F \Bigr ) + c(\mu) \, \mathbb 1 \, ,
\ee
where
\be
\omega_{k i}^B=
\begin{cases} 
\displaystyle + \sqrt{k^2+b_i^2\mu^2} \, , & k \ge 0 \\[1ex]
\displaystyle -\sqrt{k^2+b_i^2\mu^2} \, , & k < 0 
\end{cases}\, , \quad 
\omega_{k i}^F=
\begin{cases}
\displaystyle + \sqrt{k^2+f_i^2\mu^2} \, , & k \ge 0 \\[1ex]
\displaystyle -\sqrt{k^2+f_i^2\mu^2} \, , & k < 0
\end{cases} \, ,
\ee
and
\be
N_{k i}^B=
\begin{cases}
\displaystyle \frac{1}{\omega_{k i}^B} \alpha^i_{-k} \alpha^i_k \, , & k>0  \\[2.5ex]
\displaystyle {a^i}^\dagger a^i \, , & k = 0 \\
\displaystyle \frac{1}{\omega_{-k i}^B} \widetilde\alpha^i_k \widetilde\alpha^i_{-k} \, , & k<0
\end{cases} \, , \quad 
N_{k i}^F=
\begin{cases}
\displaystyle S^i_{-k} S^i_k \, , & k>0  \\[1ex]
\displaystyle {s^i}^\dagger s^i \, , & k = 0 \\[1ex] \widetilde S^i_k \widetilde S^i_{-k} \, , & k<0
\end{cases} \, .
\ee
Finally, $c(\mu)$ denotes the normal order constant
\be
c(\mu) = \frac12 \sum_{k \in \mathds Z} \sum_{i=1}^8 \( |\omega_{k i}^B| - |\omega_{k i}^F| \) \, .
\ee
Let us stress that $c$ goes to zero as $\mu\to0$.

Notice that, given
\begin{subequations}\label{brackets}
\begin{gather}
\begin{align}
&\[\alpha_k^i, \alpha_l^j\]=\[\widetilde \alpha_k^i, \widetilde \alpha_l^j\] =  \delta^{ij} \delta_{k+l,0} \, \omega_{k i}^B \, , \quad \[a^i, {a^j}^\dagger\]= \delta^{ij} \, ,\\
&\{S_k^i, S_l^{j}\} = \{\widetilde S_k^i, \widetilde S_l^{j}\} = \delta^{ij} \delta_{k+l,0} \, , \quad \{s^i, {s^j}^\dagger\}= \delta^{ij} \, ,\\
&\label{vac}\,\,\alpha^i_k | 0 \rangle = \widetilde \alpha^i_k | 0 \rangle = S^i_k | 0 \rangle = \widetilde S^i_k | 0 \rangle = 0 \, , \quad \forall k>0 \, , \quad a^i | 0 \rangle = s^i | 0 \rangle = 0 \, ,
\end{align}
\end{gather}
\end{subequations}
it is easy to realize that
\begin{subequations}\label{brackets}
\begin{gather}
\begin{align}
&  N_{k i}^B \( \beta_{-k}^i \)^{\hspace{-1pt}n_{k i}^B} | 0 \rangle = n_{k i}^B  \( \beta_{-k}^i \)^{\hspace{-1pt}n_{k i}^B} | 0 \rangle \, , \quad n_{k i}^B = 0, 1, ... , +\infty \, , \quad k \in \mathds Z \, ,\\
& N_{k i}^F \( \sigma_{-k}^i \)^{\hspace{-1pt}n_{k i}^F} | 0 \rangle = n_{k i}^F  \( \sigma_{-k}^i \)^{\hspace{-1pt} n_{k i}^F} | 0 \rangle \, , \quad n_{k i}^F = 0, 1 \, , \quad k \in \mathds Z \, , 
\end{align}
\end{gather}
\end{subequations}
where
\be
\beta_{-k}^i=
\begin{cases}
\displaystyle
\alpha^i_{-k} \, , & k>0  \\
\displaystyle {a^i}^\dagger \, , & k = 0 \\ \displaystyle \widetilde\alpha^i_{k} \, , & k<0
\end{cases} \, , \quad 
\sigma_{-k}^i=
\begin{cases} 
\displaystyle S^i_{-k} \, , & k>0  \\ {s^i}^\dagger \, , & k = 0 \\
\displaystyle \widetilde S^i_k \, , & k<0 \end{cases} \, .
\ee

In flat space, the levels of the string spectrum are labeled by the (integer) eigenvalues $\mathfrak n$ of the total number operator
\be \label{totalNflat}
\mathcal{N} = \sum_{k \in \mathds Z} \sum_{i=1}^8 |k| \Bigl [ N_{k i}^B + N_{k i}^F \Bigr ] \, , \quad \mu = 0 \, ,
\ee
that is
\be
\mathfrak n = \sum_{k \in \mathds Z} \sum_{i=1}^8 |k| \Bigl [ n_{k i}^B + n_{k i}^F \Bigr ] \, .
\ee
Notice that the operator in \eqref{totalNflat} is the $\mu\to0$ limit of $\mathcal H$ in \eqref{Hppwave}.

So, it is very convenient to consider a quantity like
\be \label{tracewN}
\Tr_{\text{phys}} \hspace{-2pt} \(w^{\mathcal N}\) = \sum_{\mathfrak n} d(\mathfrak n) \, w^{\mathfrak n} \, ,
\ee
where the trace runs over the set of physical stringy states, $w$ is a real auxiliary variable\footnote{In principle, $w$ has nothing to do with the temperature of the system. Notice that the convergence of the series in \eqref{tracewN} requires $w<1$. Indeed, it is usually parameterized as $w=e^{-\beta}$, being $\beta$ a positive quantity (e.g., see \cite{Russo:2002rq}).} and $d(\mathfrak n)$ is the degeneracy of the $\mathfrak n$th-level. Clearly, the latter is the density of
states we are looking for and can be obtained as the coefficient of the $\mathcal{O}(w^{\mathfrak n})$-term in the Taylor expansion of $\Tr_{\text{phys}}\hspace{-2pt} \(w^{\mathcal N}\)$ around $w=0$. Therefore, \eqref{tracewN} is known as the generating function of the density of states.

From the well-known Cauchy's integral formula, it follows that
\be \label{densitydefinition}
d(\mathfrak n) = \frac{1}{2 \pi i} \oint \frac{\rmd z}{z^{\mathfrak n+1}} \Tr_{\text{phys}} \hspace{-2pt} \(z^{\mathcal N}\) \, .
\ee
Notice that, as it will be clear in the following,\footnote{The generating function turns out to be an infinite product of terms that go to one as $w$ goes to zero. Then, in this limit, we get what we are claiming. Furthermore, we have not taken any high energy limit yet. Therefore, it is perfectly reasonable to consider what is the result for $\mathfrak n = 0$.} with this notation we have $d(0)=1$. In other words, we are omitting the degeneracy of the ground state.

In a curved background, $\mathcal{H}$ has non-integer real eigenvalues
\be
\mathcal{E} = \sum_{k \in \mathds Z} \sum_{i=1}^8 \Bigl [|\omega_{k i}^B| \, n_{k i}^B + |\omega_{k i}^F| \, n_{k i}^F \Bigr ]  + \, c(\mu)\, .
\ee
Let us stress that the mass-shell condition of the model reduces to
\be \label{asymmassshell}
\mathcal E = \alpha' M^2/2
\ee
on a generic stringy state, $M^2$ being its mass squared. Notice that a concept of mass can be defined whenever the background includes a Minkowskian sector. More precisely, it corresponds to the modulus squared of the canonical momentum of the string in those directions. The latter can be identified with its center-of-mass momentum. Since we are dealing with a Lorentzian background, such sector exists and has dimension at least equal to one. 
 
As in \cite{Russo:2002rq}, we consider the 
generalization of \eqref{densitydefinition} 
to curved space as
\be \label{curveddensitydefinition}
d(\mathcal{E}; \mu) = \frac{1}{2 \pi i} \oint \frac{\rmd z}{z^{\mathcal{E}+1}} \, \Tr_{\text{phys}} \hspace{-2pt} \(z^\mathcal{H}\) \, .
\ee
What we have to compute is thus the 
generating function appearing in the above line integral. 
Notice that the translational invariance along the Minkowskian (flat) directions of the target space induces a degeneracy of the vacuum state in \eqref{vac}. Indeed, we should have equipped such a state with a label $\vec p$ so that $\left | 0, \vec p \, \right \rangle$ represents a non-vibrating 
string with center-of-mass (spatial) momentum $\vec p$
 . If the target space is curved, then the integral in \eqref{curveddensitydefinition} can depend on the momenta through the mass-scale $\mu$ of the world-sheet modes. For instance, as we will see, it happens in the pp-wave scenario, where $\mu$ is proportional to the light-cone momentum. Therefore, here we are focusing on the internal degrees of freedom of a single string at fixed~$\vec p$.

First of all, let us enlarge the trace $\Tr_{\text{phys}}$ including also all the possible non-physical states, that is the states not satisfying the level matching condition. This can be achieved imposing the level matching constraint through a Lagrange multiplier $\varphi$ as
\be \label{lvlmatchmoltppwave}
\Tr_{\text{phys}}\(w^\mathcal{H}\) = \frac{1}{2\pi} \int_0^{2\pi} \hspace{-6pt} \rmd\varphi \, \, \Tr \( w^\mathcal{H} e^{i \varphi \mathcal{P}} \) \, , 
\ee
where
\be \label{Pppwave}
\mathcal{P} = \sum_{k \in \mathds Z} \sum_{i=1}^8 \, k \Bigl [N_{k i}^B + N_{k i}^F \Bigr ] 
\ee
is the world-sheet momentum and the trace $\Tr$ runs over all the possible states like
\vspace{-0.5pt}
\be \label{possiblestate}
\prod_{k \in \mathds Z} \prod_{i}^8 \( \beta_{-k}^i \)^{\hspace{-1pt}n_{k i}^B} \( \sigma_{-k}^i \)^{\hspace{-1pt}n_{k i}^F} | 0 \rangle \, ,
\ee
for any assignment of the set of the occupation numbers $\{n_{k i}^B\}_{k\in\mathds Z}^{i=1,...,8}$ and $\{n_{k i}^F\}_{k\in\mathds Z}^{i=1,...,8}$.\footnote{In \eqref{Hppwave} we adopt a compact notation formally including also the contributions of $N_{0 i}^{B,F}$ for $b_i$, $f_i=0$. Nevertheless, they are set to zero since $\omega_{0 i}^{B,F}=0$ if the mass of the corresponding mode is zero. Therefore, with an abuse of notation, we tacitly do not include $n_{0 i}^{B,F}$ in the set $\{n_{k i}^{B,F}\}_{k\in\mathds Z}^{p=1,...,8}$ for all $i$ such that $b_i$, $f_i=0$. Roughly speaking, the on-shell massless fluctuations do not involve zero-modes.} 
 
Given this notation, we can write
\be \label{nonphystrace}
\Tr \( w^\mathcal{H} e^{i \varphi \mathcal{P}} \) = e^{c(\mu)\log w} \hspace{-4pt}\prod_{\mathcal{S}=B,F} \sum_{\{n_{k i}^\mathcal{S}\}_{k\in\mathds Z}^{i=1,...,8}} \, \prod_{p\in\mathds Z} \,  \prod_{j=1}^8  e^{\( |\omega_{p j}^\mathcal{S}| \log w + i \varphi \, p \) n_{p j}^\mathcal{S} } \, .
\ee
Notice that the dependence on the occupation numbers is factorized. Moreover, each occupation number takes values from a domain which is not affected by the other ones. So, it is easy to realize that
\vspace{-5pt}
\be
\Tr \( w^\mathcal{H} e^{i \varphi \mathcal{P}} \) = e^{c(\mu)\log w} \hspace{-4pt}\prod_{\mathcal{S}=B,F} \, \prod_{\substack{k\in\mathds Z\\ k \neq 0} } \, \prod_{\substack{i\\\omega_{0 i}^\mathcal{S} \neq0}}^8 \, \prod_{j=1}^8 \,  \sum_{n_{0i}^\mathcal{S}=0}^{\mathcal{U_S}} \, \sum_{n_{k j}^\mathcal{S}=0}^{\mathcal{U_S}} e^{\omega_{0 i}^\mathcal{S} \log w \, n_{0i}^\mathcal{S}} \, e^{\(|\omega_{k j}^\mathcal{S}| \log w + i \varphi k \) n_{k j}^\mathcal{S}}  \, ,
\ee
where $\mathcal{U}_B=+\infty$, $\mathcal{U}_F=1$.

Clearly, the sums over $n_{0i}^B$ and $n_{kj}^B$ are just geometric series. Therefore, we get
\begin{align} \label{finalppwavetrace}
\Tr \( w^\mathcal{H} e^{i \varphi \mathcal{P}} \) &= e^{c(\mu)\log w} \prod_{i=1}^8 \, \prod_{k \in \mathds Z} \, \frac{1+e^{\omega_{ki}^F \log w + i \varphi k}}{1-e^{\omega_{ki}^B \log w + i \varphi k}} \\
&= e^{c(\mu)\log w} \hspace{-4pt} \prod_{\substack{i\\b_i, f_i \neq0}}^8 \hspace{-4pt} \frac{1+w^{f_i\mu}}{1-w^{b_i\mu}} \, \prod_{k=1}^\infty \, \prod_{j=1}^8 \, \left | \frac{1 + e^{\omega_{k j}^F \log w + i \varphi k}}{1-e^{\omega_{kj}^B \log w + i \varphi k}} \right |^2 \, ,
\end{align}
whence the density of states takes the form
\be \label{dppwavestart}
d(\mathcal{E}; \mu) = \frac{1}{2 \pi i} \oint \frac{\rmd z}{z^{\mathcal{E}+1}} \, e^{c(\mu)\log w} \, \frac{1}{2\pi}\int_0^{2\pi} \hspace{-6pt} \rmd\varphi \hspace{-4pt} \prod_{\substack{i=1\\b_i, f_i \neq0}}^8 \hspace{-4pt} \frac{1+z^{f_i\mu}}{1-z^{b_i\mu}} \, \prod_{k=1}^\infty \, \prod_{j=1}^8 \, \left | \frac{1 + e^{\omega_{k j}^F \log z + i \varphi k}}{1-e^{\omega_{kj}^B \log z + i \varphi k}} \right |^2 \, .
\ee
This is very reminiscent of formulae (6.2), (6.4) and (6.5) of \cite{Russo:2002rq}. 

Now, let us suppose that $10-D$ of the spatial coordinates are compactified into a torus. So far, we took into account just the contributions to the density of states coming from the string oscillators. Let us discuss how to include also the Kaluza-Klein and the winding modes associated with the compactification.

The background metric can be decomposed as
\be \label{genback}
\rmd s^2 =\rmd s^2_{\text{non-compact}} + \rmd s^2_{T^{10-D}} \, .
\ee
It follows that the oscillatory part of the canonical Hamiltonian and the world-sheet momentum must be mapped to
\begin{subequations}
\begin{align}
&\label{shiftedH}\mathcal H \mapsto \mathcal H + \frac12 \, \sum_{j=1}^{10-D} \[\(m_j \frac{\sqrt{\alpha'}}{R_j}\)^2 + \(w_j \frac{R_j}{\sqrt{\alpha'}}\)^2\] \, ,\\
&\mathcal P \mapsto \mathcal P + \sum_{j=1}^{10-D} m_j \, w_j \, .
\end{align}
\end{subequations}
Here, $m_j$ and $w_j$ are both integer numbers for all $j=1,...,10-D$. They respectively label the quantized momentum and the winding number along the $j$-th direction. The latter is characterized by a compactification radius $R_j$. Accordingly, $\mathcal E$ is now the eigenvalue of the whole shifted operator in \eqref{shiftedH}, so that the mass-shell condition still realizes in \eqref{asymmassshell}.

We already have all we need to extend the previous result in the presence of $10-D$ compact directions. In particular, the trace over the set of non-physical states in \eqref{nonphystrace} gets modified as
\be
\Tr \(w^{\mathcal H} e^{i \, \varphi \, \mathcal P}\) \mapsto \mathcal K (\varphi, w) \, \Tr \(w^{\mathcal H} e^{i \, \varphi \, \mathcal P}\) \, ,
\ee
where
\begin{align}
\mathcal K (\varphi, w) &= \sum_{\{w_j,m_j\}_{j=1}^{10-D}} \prod_{k=1}^{10-D} \(w^{\frac{R_k^2}{2\,\alpha'}}\)^{\hspace{-3pt}w_k^2} \(w^{\frac{\alpha'}{2\,R_k^2}}\)^{\hspace{-3pt}m_k^2} e^{i \, \, \varphi \, m_k w_k} \nonumber \\
&= \prod_{k=1}^{10-D} \sum_{w_k,m_k \in \mathds Z}  \(w^{ \frac{R_k^2}{2\,\alpha'}}\)^{\hspace{-3pt}w_k^2} \(w^{ \frac{\alpha'}{2\,R_k^2}}\)^{\hspace{-3pt}m_k^2} e^{i \, \varphi \, m_k w_k} \nonumber \\
& = \prod_{k=1}^{10-D} \sum_{w_k,m_k \in \mathds Z}  \(w^{ \frac{R_k^2}{2\,\alpha'}}\)^{\hspace{-3pt}w_k^2} \(w^{ \frac{\alpha'}{2\,R_k^2}}\)^{\hspace{-3pt}m_k^2} \cos\( \varphi \, m_k w_k\) \, .
\end{align}
All in all, the final result for the density of states is given by
\be \label{dppwavestartcompact}
d(\mathcal{E}; \mu) = \frac{1}{2 \pi i} \oint \frac{\rmd z}{z^{\mathcal{E}+1}} \, e^{c(\mu)\log w} \, \frac{1}{2\pi}\int_0^{2\pi} \hspace{-6pt} \rmd\varphi \,\, \mathcal K (\varphi, z) \hspace{-2pt} \prod_{\substack{i=1\\b_i, f_i \neq0}}^8 \hspace{-4pt} \frac{1+z^{f_i\mu}}{1-z^{b_i\mu}} \, \prod_{k=1}^\infty \, \prod_{j=1}^8 \, \left | \frac{1 + e^{\omega_{k j}^F \log z + i \varphi k}}{1-e^{\omega_{kj}^B \log z + i \varphi k}} \right |^2 \, .
\ee
The latter is valid in any regime for a solvable closed superstring model on a generically curved background featuring $D$ non-compact directions. Let us stress that $d(\mathcal E; \mu) \rmd \mathcal E$ counts how many states in the single-string spectrum have ``energy" between $\mathcal E$ and $\mathcal E + \rmd \mathcal E$ at fixed center-of-mass momentum $\vec p$.

\section{The flat space case: a brief review}
\label{sec:flat}

In this work we aim to discuss the asymptotic behavior of \eqref{dppwavestart} and \eqref{dppwavestartcompact} for very highly excited states. Let us try to get some insights from well-known cases. In particular, let us start the analysis dealing with bosonic open strings and superstrings in flat space.

\subsection{Bosonic open string in flat space}
\label{sec:bosopstring}

The density of states for an open bosonic string embedded in flat space is given by (see, e.g., \cite{Green:2012oqa})
\be \label{dnflatstart}
g(n) = \frac{1}{2 \pi i} \oint \frac{\rmd z}{z^{n+1}} \, G(z) \, , \quad G(z) = \prod_{k=1}^\infty \frac{1}{(1-z^k)^{24}} \, ,
\ee
$n$ being the eigenvalue of the number operator.\footnote{Let us stress that it is different from $\mathfrak n$. The latter is the total number of (right and left) closed string oscillators introduced in section \ref{sec:genform}.}

It is well known that the integrand displays a sharp saddle point at $z\sim1$ in the large-$n$ limit.\footnote{Roughly speaking, due to the presence of the factor $z^{-n}=\text{exp}(-n\log z)$, in the large-$n$ limit the final result is expected to be ruled by the neighborhood of $z=1$ in the domain of integration.} As a consequence, the asymptotic behavior of $g(n)$ is encoded in the expansion of $G(z)$ around $z=1$, that is
\be
G(z) \sim \text{(const.)} (-\log z)^{12} e^{-4 \pi^2/\log z} \, .
\ee

In other words, in the large-$n$ regime, $g(n)$ is approximately given by the integral of $G(z)/z^{n+1}$ along a small path which passes by a point approaching $z=1$ as $n \to +\infty$. Let us parameterize this curve in the complex $z$-plane as the circumference
\be \label{zpath}
z(\alpha) = \rho \, e^{i \alpha} \, , \quad \rho~\text{constant}, \, \quad \alpha \in \(-\pi, +\pi\] \, .
\ee
We can thus estimate \eqref{dnflatstart} as
\be\label{dnflatapprox}
g(n) \sim \frac{1}{2 \pi} \int_{-\varepsilon}^{+\varepsilon} \hspace{-8pt} \rmd\alpha \, f(\alpha) \, e^{\Phi(\alpha)} e^{i \, \Psi(\alpha)} \, ,
\ee
once we take $\rho \sim 1$. Here, $\varepsilon$ is a small positive real number and
\be
\Phi(\alpha)=\text{Re}\{-\frac{4 \pi^2}{\log z(\alpha)} - n \log z(\alpha)\} \, , \quad \Psi(\alpha)=\text{Im}\{-\frac{4 \pi^2}{\log z(\alpha)} - n \log z(\alpha)\} \, ,
\ee
that is
\be\label{Phi}
\Phi(\alpha) = \[\frac{4\pi^2}{(\log \rho)^2 + \alpha^2} + n\] (-\log \rho)
\ee
and
\be
\Psi(\alpha) = \[\frac{4\pi^2}{(\log \rho)^2 + \alpha^2} - n\] \alpha \, .
\ee
Moreover,\footnote{In a neighborhood of $\rho\sim1$, the logarithm of $\rho$ can be slightly negative or positive. In the former case, we have that $\arg(-\log \rho-i \, \alpha)=\arctan(\alpha/\log \rho)$. In the second one, we have $\arg(-\log \rho-i \, \alpha)=\arctan(\alpha/\log \rho) \pm \pi$. Since the argument function here is multiplied by 12 and it is meant to be the phase of a plane wave, we can just take $\arg(-\log \rho-i \, \alpha)=\arctan(\alpha/\log \rho)$.}
\be
f(\alpha) = (-\log z(\alpha))^{12} = \[(\log \rho)^2 + \alpha^2\]^6 \, e^{ 12 \, i \arctan\[\frac{\alpha}{\log \rho}\]}.
\ee

The idea is to evaluate \eqref{dnflatapprox} by means of the method of steepest descent. So we have to fix the path of integration in such a way that $\Phi(\alpha)$ displays a sharp maximum in a neighborhood of $\alpha = 0$ (namely, $\alpha \in \[-\varepsilon,+\varepsilon\]$), while $\Psi(\alpha)$ is constant. In other words, we have to understand how to fix $\rho$, which is currently the only free parameter, in order to meet the just mentioned requirements.

With
\be
\left . \frac{\partial \Phi(\alpha)}{\partial \alpha} \right |_{\alpha=0} = 0 \, , \quad \left . \frac{\partial \Psi(\alpha)}{\partial \alpha} \right |_{\alpha=0} = - n + \frac{4 \pi^2}{(\log\rho)^2} \, ,
\ee
both $\Phi$ and $\Psi$ are stationary in $\alpha=0$ if
\be
\rho = e^{\pm \frac{2\pi}{\sqrt{n}}} \,.
\ee
This corresponds to a stationary point in $z \sim 1$ in the large-$n$ limit such that
\be
\left . \frac{\partial^2 \Phi(\alpha)}{\partial \alpha^2} \right |_{\alpha=0} = \frac{8\pi^2}{(\log\rho)^3} = \pm \frac{n^{3/2}}{\pi} \, , \quad \left . \frac{\partial^2 \Psi(\alpha)}{\partial \alpha^2} \right |_{\alpha=0} = 0 \, .
\ee

\begin{figure}[t]
	\scalebox{0.32}{\includegraphics{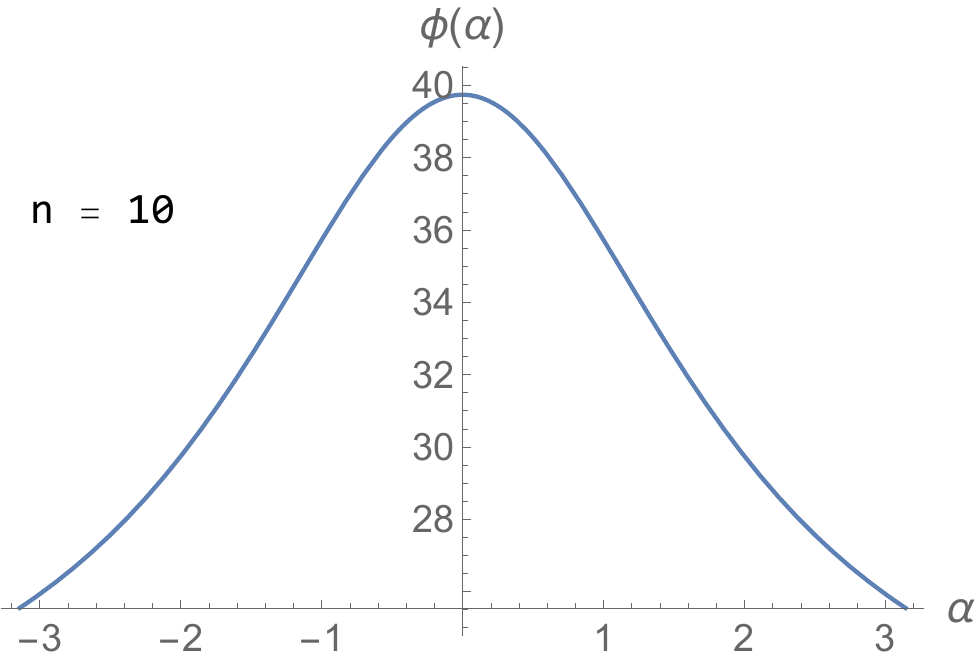}}\,\,\,
	\scalebox{0.32}{\includegraphics{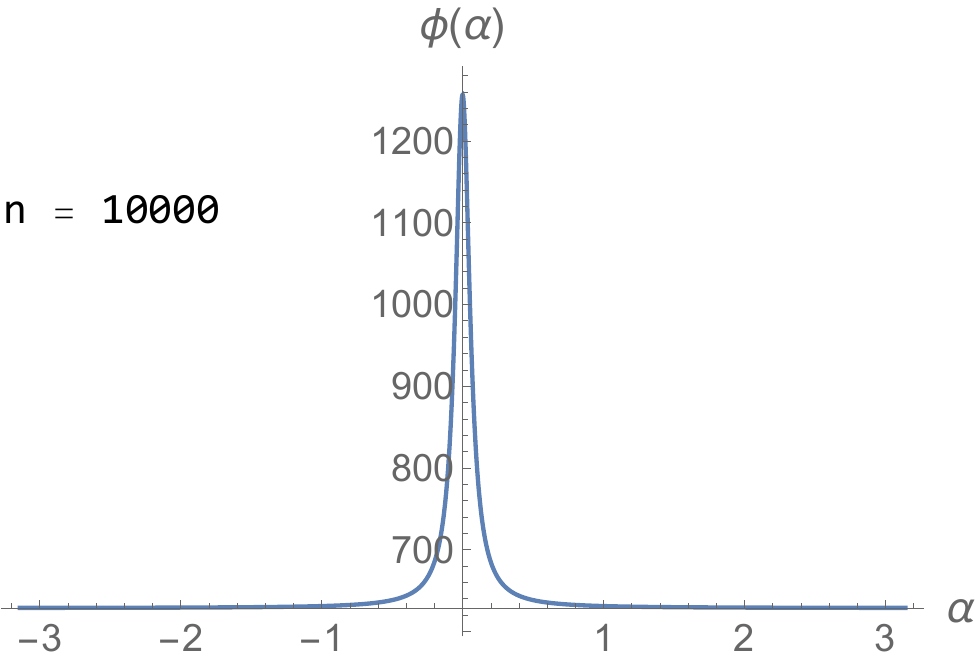}}\,\,\,
	\scalebox{0.32}{\includegraphics{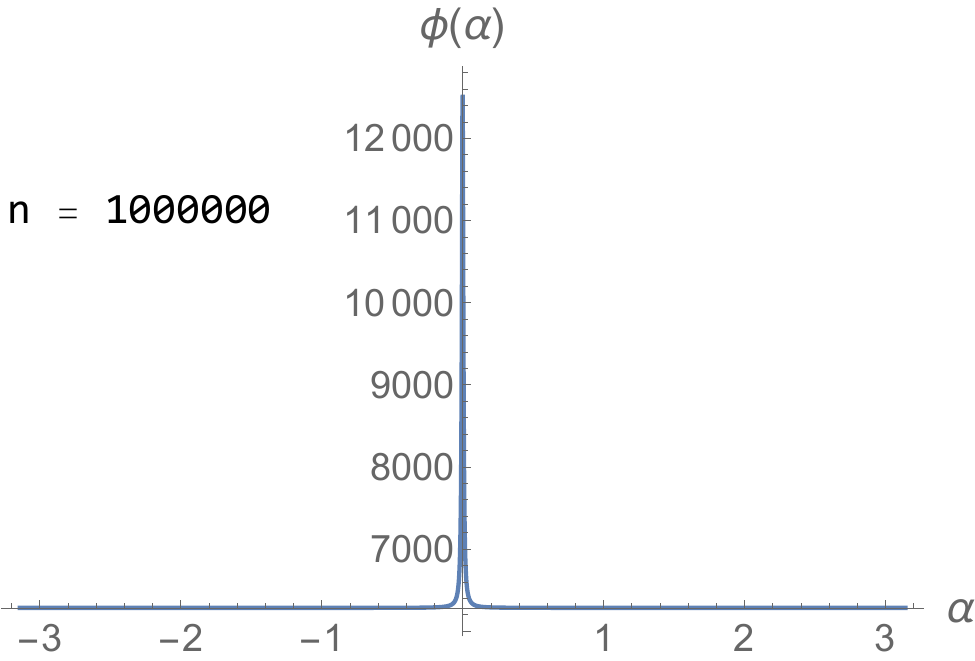}}
	\caption{Plots of $\Phi=\Phi(\alpha)$ in \eqref{Phi} for $\log\rho=-2\pi/\sqrt{n}$ at different values of $n$.}
	\label{fig:sharpness}
\end{figure}
Then, we conclude that, given $\rho = e^{- \frac{2\pi}{\sqrt{n}}}$, $\alpha=0$ is a maximum point of $\Phi$ around which $\Psi$ is constant. The larger $n$ is, the sharper the maximum is (see plot \ref{fig:sharpness}). All in all, with
\be
\Phi(0) = 4\pi\sqrt{n} \, , \quad \Psi(0) =0 \, , \quad f(0) = \(\frac{2\pi}{\sqrt{n}}\)^{\hspace{-3pt}12} \, ,
\ee
we can approximate \eqref{dnflatapprox} as
\be
g(n) \sim \frac{1}{2 \pi} \, e^{4\pi\sqrt{n}} \(\frac{2\pi}{\sqrt{n}}\)^{\hspace{-3pt}12} \int_{-\infty}^{+\infty} \hspace{-4pt} \rmd\alpha \, \,e^{-\frac12 \frac{n^{3/2}}{\pi} \alpha^2} \,  \, , \quad n \to +\infty \, ,
\ee
that is
\be
g(n) \approx 
n^{-\frac{27}{4}} e^{4\pi\sqrt{n}}  \,  \, , \quad n \to +\infty \, .
\ee
This is exactly the well-known result we expected to find \cite{Green:2012oqa}.

\subsection{Superstring in flat space}
\label{sec:superflat}

Now, let us take a step further introducing the fermionic degrees of freedom. The density of states for an open superstring in flat space reads (see, e.g., \cite{Sundborg:1984uk})
\be \label{dnflatsuper}
h(n) = \frac{1}{2 \pi i} \oint \frac{\rmd z}{z^{n+1}} \, \frac{1}{\theta_4^8(0,z)} \, , \quad \frac{1}{\theta_4^8(0,z)} = \prod_{k=1}^\infty\(\frac{1+z^k}{1-z^k}\)^{\hspace{-4pt}8} \, ,
\ee
where the above Jacobi Theta Function $\theta_4$ is defined in appendix \ref{app:theta}. %
As in the previous section, the integrand displays a saddle point at $z\sim1$.
Given the asymptotic expression in \eqref{asymtheta4}, %
we can follow the same steps as before. The final result is
\be
h(n) \approx n^{-\frac{11}{4}} e^{\pi\sqrt{8n}} \, , \quad n \to +\infty \, .
\ee

The density of states for a closed superstring in flat space can be derived in a moment from the above expression. Indeed, in flat space, the oscillatory part of the string Hamiltonian contains the same number operators appearing in the world-sheet momentum $\mathcal{P}$ (cfr.~\eqref{totalNflat}, \eqref{Pppwave}).
Therefore, the level matching condition can be imposed at the end of the computation, dealing with left and right oscillators independently (see footnote 19 of \cite{Russo:2002rq}). In other words,
\be \label{dclflat}
d_{\text{flat}}(n) = (h(n))^2 \approx n^{-\frac{11}{2}} e^{2\pi\sqrt{8n}} \, , \quad n \to +\infty \, .
\ee
Here, $n$ stands for the eigenvalue of the left or right number operator, which are constrained to be equal to each other by the level matching condition.

Crucially, no one forbids to impose the level matching condition by integrating over a Lagrange multiplier $\varphi$ as in \eqref{lvlmatchmoltppwave}, working with a number of left movers $n$ which is in principle different from the number of right movers $\tilde n$. Thus, formally, the density of state can also be written as
\be \label{dclflatlambda}
d_{\text{flat}}(\mathcal{E}) = \frac{1}{2 \pi i} \oint \frac{\rmd z}{z^{\mathcal{E}+1}} \, \frac{1}{2\pi} \int_{0}^{2\pi} \hspace{-6pt} \rmd\varphi \,\, \Pi(\varphi,z) \, , \quad \mathcal{E}=n+\tilde n \, .
\ee
The above formula is the flat-space analogue of \eqref{dppwavestart}. In particular, $\Pi$ can be derived following the steps of section \ref{sec:genform}. Basically, we can read the final result directly from \eqref{finalppwavetrace}, ignoring the zero mode contributions and fixing $\mu=0$.
It follows that
\be \label{Pi}
\Pi(\varphi, w) = \prod_{k=1}^{\infty} \, \left | \frac{1 + w^k e^{+ i \varphi k}}{1-w^k e^{+ i \varphi k}} \right |^{16} \, .
\ee

Of course, the density of states in \eqref{dclflatlambda} has to be in agreement with \eqref{dclflat}, that is
\be \label{dclflatlambdaexplicit}
d_{\text{flat}} (\mathcal E) \approx \mathcal E^{-\frac{11}{2}} \, e^{2\pi\sqrt{4\,\mathcal E}} \, , \quad \mathcal E \to + \infty \, ,
\ee
where at the end of the story $\mathcal E = 2 n$. This is exactly the case. Indeed, notice that the $z$-integrand in \eqref{dclflatlambda} displays a saddle point at $z \sim 1$ as usual.\footnote{The large parameter is $\mathcal{E}$. Therefore, due to the factor $z^{-\mathcal{E}}=e^{-\mathcal{E} \log z}$, the non-trivial contribution to the final result is again expected to come from the $z\sim1$ region of the domain of integration.} 
Referring to the formulae of appendix \ref{app:theta} for the Jacobi Theta Functions, 
it is straightforward to check that
\be \label{ansatz}
\frac{1}{2\pi} \int_{0}^{2\pi} \hspace{-4pt} \rmd\varphi \,\, \Pi(\varphi,w) \sim \frac14 \, \theta_4^{-16}(0,w) \, \theta_2^{-3}(0,w) \, , \quad w \to 1^- \, .
\ee
Let us stress that the details about the full result of the $\varphi$-integration are not relevant. What matters is just the behavior around $z\sim1$, 
which produces the expected outcome for $d_{\text{flat}}$. See appendix \ref{app:Pi} for both an analytical computation and a numerical check of the above formula.

\subsection{Toroidal compactification}
\label{sec:torcom}

As a final step, let us compactify $10-D$ of the flat directions into a torus. From the discussion at the end of section \ref{sec:genform}, it is easy to realize that all we need to do is mapping $\Pi$ in \eqref{Pi} as
\be \label{Pishift}
\Pi(\varphi,w) \mapsto \mathcal K (\varphi, w) \, \Pi\(\varphi,w\) \, .
\ee

In appendix \ref{app:Pi} we show that
\be \label{ansatzcompact}
\frac{1}{2\pi} \hspace{-2pt} \int_{0}^{2\pi} \hspace{-10pt} \rmd\varphi \,\, \mathcal K (\varphi, w) \, \Pi(\varphi,w) \sim \frac{\prod_{k=1}^{10-D} \theta_3\hspace{-2pt}\(0, w^{\frac{R_k^2}{2\,\alpha'}}\) \theta_3\hspace{-2pt}\(0,w^{\frac{\alpha'}{2\,R_k^2}}\)}{4 \, \theta_4^{16}(0,w) \, \theta_2^{3}(0,w)} \, , \quad w \to 1^- \, .
\ee
With this result at hand, we can compute the density of states similarly to the previous section as
\be \label{dcflattorus}
d_{\text{flat$\times$torus}}\(\mathcal E\) \approx \mathcal{E}^{-\frac{D+1}{2}} \, e^{4\pi\sqrt{\mathcal{E}}} \, , \quad \mathcal E \to +\infty \, .
\ee
Notice that, for $D=10$, it reduces to the flat space density of states in \eqref{dclflatlambdaexplicit}.

This result reproduces the outcome in \cite{Dienes:1998hx} for a Type II string in flat space compactified to $D$ dimensions. There, the authors derived it by neglecting the presence of the compact directions and compensating for them by including their effects a posteriori. They concluded that only the power law is affected
, relying just on the conformal properties of the world-sheet sigma model. This is exactly what we found here from an explicit computation. Indeed, the extra Jacobi Theta Functions coming from $\mathcal K$ accomplishes the above.

\section{Closed superstring in 
curved background}
\label{sec:ppwave}

Let us 
apply what we learned to the general case of interest \eqref{genback}. Of course, life is not so easy and we have to face some problems going through with the computation. 

First of all, in curved space, the operators contained in the oscillatory part \eqref{Hppwave} of the string Hamiltonian do not correspond to the ones appearing in the world-sheet momentum $\mathcal{P}$ in \eqref{Pppwave}. Therefore, we cannot use the trick of the previous section and we are forced to deal with a Lagrange multiplier in order to impose the level matching condition.

Moreover, the final result for the density of states in the previous cases was $\vec p$-independent. In the general case of interest, we already stressed that \eqref{dppwavestartcompact} represents the density of states for a single string at fixed center-of-mass momentum. This is crucial because we aim to apply our result to string thermodynamics. As it will be clear in the next section, we need to integrate over all the possible momenta of the single constituents to compute the partition function of a gas of strings. Therefore, $\mu$ can potentially span whatever value and so we need to rephrase the starting integral in \eqref{dppwavestartcompact} in such a way the $\mu$-dependence is under control. This is a reminiscent of the problems discussed in \cite{Hyun:2003ks}, where the authors pointed out how the behavior of the partition function in the Hagedorn regime is polluted by the largeness of the world-sheet mass-scale. To get these issues under control, some manipulations are in order.

To begin with, let us introduce
\be \label{Dfunction}
D_{b_1,\,b_2}\(\tau_1, \tau_2; m\) = e^{ 2\pi\tau_2 \,\Delta_{b_1}(m)} \prod_{k \in \mathds Z} \(1-e^{-2\pi\tau_2\sqrt{\(k+b_1\)^2 + m^2} + 2\pi i \, \tau_1 \(k + b_1\) - 2\pi i \, b_2}\) \, ,
\ee
where
\begin{align}\label{zeropointenergies}
\Delta_b (m) &= - \frac{m}{\pi} \sum_{p=1}^\infty \frac{\cos (2\pi b p)}{p} K_1\(2\pi m p \) \nonumber\\
&= \frac12 \sum_{k \in \mathds Z} \sqrt{(k+b)^2+m^2} - \frac12 \int_{-\infty}^{+\infty} \rmd k \sqrt{(k+b)^2 + m^2}\, .
\end{align}
As stressed in \cite{Hyun:2003ks}, $\Delta_b$ corresponds to the regularized form of the zero point (or Casimir) energy for two-dimensional massive bosonic ($b=0$) or fermionic ($b=1/2$) fields having twisted boundary condition and mass $m$. Moreover, it admits a power series expansion in terms of $m$ as
\be \label{Casimirexpansion}
\begin{cases}
\begin{split}
 \Delta_0(m) = - \frac{1}{12} + \frac12 m + &\frac14 m^2 \[\log\frac{m^2}{4} + 2 \, \gamma_E - 1\] + \\& + \sum_{k=2}^{\infty} \frac{(-1)^k \Gamma\(k - \frac12\)}{\Gamma \(-\frac12\) \Gamma \(k+1\)}  \zeta\(2 k -1\) m^{2k}
\end{split}\, ,\\[2ex]
\begin{split}
\Delta_{\sfrac12}(m) = \frac{1}{24} + &\frac14 m^2 \[\log\frac{m^2}{4} + 2 \, \gamma_E - 1 + 4 \log 2\] +\\
&+ \sum_{k=2}^{\infty} \frac{(-1)^k \Gamma\(k - \frac12\)}{\Gamma \(-\frac12\) \Gamma \(k+1\)} (2^{2 k -1} - 1) \zeta\(2 k -1\)m^{2k}
\end{split}\, .
\end{cases}
\ee
Here, $\gamma_E=-\left . \rmd \log \Gamma(q)/\rmd q \right |_{q=1} = 0.5772 \ldots$ is the Euler constant.

It thus follows that the trace in \eqref{finalppwavetrace} equates the function
\be
\mathcal G (\varphi, w) = C_0 (w) \, \prod_{i=1}^{\text N_b^{\vphantom{k}}} \prod_{j=1}^{\text N_f} \frac{D_{0,1/2}\(\varphi/2\pi, -\log w/2\pi; f_j \mu\)}{D_{0,1/2}\(\varphi/2\pi, -\log w/2\pi; b_i \mu\)} \, \prod_{k=1}^{\infty} \frac{\displaystyle \left |1+w^k e^{i\varphi k} \right |^{2(8- N_f)}}{\displaystyle \left |1-w^k e^{i\varphi k} \right |^{2(8- N_b)}} \, .
\ee
For the sake of simplicity, we reordered the products so that just the first $\text N_b^{\vphantom{k}}$ ($\text N_f$) bosonic (fermionic) masses are non-zero. Moreover, we defined
\be \label{C0}
C_0(w) = e^{\frac{1}{12} \(\text N_b^{\vphantom{k}}-\text N_f\) \log w }e^{\frac12 \log w \sum_{i=i}^8 \int_{-\infty}^{+\infty} \hspace{-2pt}\rmd k \,\(\sqrt{k^2 + b_i^2 \mu^2} - \sqrt{k^2 + f_i^2 \mu^2}\) } \, .
\ee
Notice that $C_0$ is finite thanks to the mass-matching condition in \eqref{massmatchppwave}. Indeed, it follows that
\be \label{subtraction}
\sum_{i=i}^8 \int_{-\infty}^{+\infty} \hspace{-6pt}\rmd k \,\(\sqrt{k^2 + b_i^2 \mu^2} - \sqrt{k^2 + f_i^2 \mu^2}\) = - \frac12 \(\sum_{i=1}^{\text N_b^{\vphantom{k}}} b_i^2 \log b_i^2 - \sum_{j=1}^{\text N_f^{\vphantom{k}}} f_j^2 \log f_j^2\) \hspace{-2pt} \mu^2 \, .
\ee
Let us stress that this contribution is trivial if the string model features the same number of world-sheet bosons and fermions with the same masses. Otherwise, it is a crucial and finite term which affects the final result.

Using the identities \cite{Hyun:2003ks}
\begin{align} \label{modidentities}
&D_{b_1,\,b_2}\(\tau_1, \tau_2; m\) = D_{b_2,\,-b_1}\(-\frac{\tau_1}{|\tau|^2}, \frac{\tau_2}{|\tau|^2}; m |\tau|\) \, , \\
&\tau = \tau_1 + i \, \tau_2 \, , \quad \(\text{for } b_1, b_2 = 0,1/2 \) \, ,
\end{align}
we can rephrase $\mathcal G$ as
\be
\mathcal G (\varphi, w) = C (\varphi, w) \, \mathcal A (\varphi, w) \left | \theta_4(0, w \, e^{i \varphi}) \right |^{2 \text N_b^{\vphantom{k}} - 16} \[\frac{\left | \theta_2(0, w \, e^{i \varphi}) \, \theta_3(0, w \, e^{i \varphi}) \right |}{2 \, |w|^{\sfrac14} \left | \theta_4(0, w \, e^{i \varphi}) \right |^2}\]^{\frac{\text N_b^{\vphantom{k}} - \text N_f}{3}} \, ,
\ee
where
\begin{subequations}
\begin{align}
& C(\varphi, w) = C_0(w) \, e^{ \frac{4 \pi^2 \log\hspace{-1pt}w}{\log^2\hspace{-1.5pt}w + \varphi^2} \(\sum_{i=1}^{\text N_b^{\vphantom{k}}} \Delta_0\(\frac{b_i \mu}{2\pi} \sqrt{\varphi^2 +  \log^2\hspace{-1.5pt}w}\) - \sum_{j=1}^{\text N_f} \Delta_{\sfrac12}\(\frac{f_j \mu}{2\pi} \sqrt{\varphi^2 +  \log^2\hspace{-1.5pt}w}\)\)} \, , \\
& \mathcal A (\varphi, w) = \prod_{i=1}^{\text N_b^{\vphantom{k}}} \prod_{j=1}^{\text N_f} \prod_{k=1}^{\infty} \prod_{r=1/2}^{\infty} \frac{1}{\sqrt{A_0 (b_i \mu)}} \frac{A_r (f_j \mu)}{A_k(b_i \mu)} \, ,
\end{align}
\end{subequations}
with
\be
A_\ell (m) = \left | 1 - e^{\frac{4\pi^2}{\log^2\hspace{-1.5pt}w + \varphi^2} \(\log w \, \sqrt{\ell^2 + m^2} - i \varphi \ell\)}\right |^2 \, .
\ee
Notice that we used the results about the Jacobi Theta functions in appendix \ref{app:theta}, so that
\be
\prod_{k=1}^{\infty} \left | \frac{1+w^k e^{i\varphi k}}{1-w^k e^{i\varphi k}} \right |^2 = \frac{1}{\left | \theta_4 (0, w \, e^{i \varphi}) \right |^2} \, , \quad \prod_{k=1}^{\infty} \left | 1+w^k e^{i\varphi k} \right |^2 = \[\frac{\left | \theta_2(0, w \, e^{i \varphi}) \, \theta_3(0, w \, e^{i \varphi}) \right |}{2 \, |w|^{\sfrac14} \left | \theta_4(0, w \, e^{i \varphi}) \right |^2}\]^{\sfrac13} \, .
\ee
See also the beginning of appendix \ref{app:Pi}.

In this way, the integral in \eqref{dppwavestartcompact} can be rewritten as
\be \label{dppwaveZlambda}
d(\mathcal{E}; \mu) = \frac{1}{2 \pi i} \oint \frac{\rmd z}{z^{\mathcal{E}+1}} \frac{1}{2\pi} \int_0^{2\pi} \hspace{-8pt} \rmd \varphi \,\, \mathcal K(\varphi,z) \, \mathcal G (\varphi,z)  \, .
\ee
Similarly to sections \ref{sec:superflat} and \ref{sec:torcom}, the integrand displays a sharp saddle point at $z\sim1$ in the large-$\mathcal E$ limit due to the presence of $z^{-\mathcal{E}}=e^{-\mathcal{E} \log z}$ in the integrand. As a consequence, we need to focus on the asymptotic form of $\mathcal G$ in this regime. 
Similarly to appendix \ref{app:Pi} (see also appendix \ref{app:theta}), we can take
\begin{subequations}
\begin{align}
&\left | \theta_4(0, w \, e^{i \varphi}) \right |^{2 \text N_b^{\vphantom{k}} - 16} \sim \(\frac{\log^{2}\hspace{-1.5pt}w + \varphi^2}{16\pi^2}\)^{4 - \frac{\text N_b^{\vphantom{k}}}{2}} e^{ \frac{\(\text N_b^{\vphantom{k}} - 8\) \pi^2 \log\hspace{-1pt}w}{2 \(\log^2\hspace{-1.5pt}w + \varphi^2\)}} \, , \quad (w \to 1^-, \varphi \to 0) \, , \\
& \[\frac{\left | \theta_2(0, w \, e^{i \varphi}) \, \theta_3(0, w \, e^{i \varphi}) \right |}{2 \, |w|^{\sfrac14} \left | \theta_4(0, w \, e^{i \varphi}) \right |^2}\]^{\frac{\text N_b^{\vphantom{k}} - \text N_f}{3}} \hspace{-4pt}\sim \, 2^{\text N_f - \text N_b^{\vphantom{k}}} \, e^{ \frac{\(\text N_f - \text N_b^{\vphantom{k}}\) \pi^2 \log\hspace{-1pt}w}{6 \(\log^2\hspace{-1.5pt}w + \varphi^2\)}} \, , \quad (w \to 1^-, \varphi \to 0) \, .
\end{align}
\end{subequations}
Moreover, $C$ can be expanded by means of \eqref{Casimirexpansion} as
\be
C(\varphi, w) = e^{ \frac{4 \pi^2 (-\log\hspace{-1pt}w)}{\log^2\hspace{-1.5pt}w + \varphi^2} \frac{\text N_f + 2 \text N_b^{\vphantom{k}}}{24}} \, \widetilde C (\varphi, w) \, ,
\ee
$\widetilde C$ being\footnote{To derive this result, we have used the mass-matching condition in \eqref{massmatchppwave}.}
\be \label{tildeC}
\widetilde C(\varphi, w) = C_0(w) \, e^{ \frac{4 \pi^2 \log\hspace{-1pt}w}{\log^2\hspace{-1.5pt}w + \varphi^2} \[\sum_{i=1}^{\text N_b^{\vphantom{k}}} \frac{b_i \mu}{4\pi}\sqrt{\log^2\hspace{-1.5pt}w + \varphi^2} - \frac{a \, \mu^2}{4\pi^2} \(\log^2\hspace{-1.5pt}w + \varphi^2\) + \sum_{k=2}^{\infty} B_k \, \frac{\mu^{2k}}{(4\pi^2)^k} \(\log^2\hspace{-1.5pt}w + \varphi^2\)^{k}\]} \, ,
\ee
where
\begin{subequations}\label{aBk}
\begin{align}
&\label{a}a = \log 2 \sum_{i=1}^{\text N_b^{\vphantom{k}}} b_i^2 - \frac14 \[\sum_{i=1}^{\text N_b^{\vphantom{k}}} b_i^2 \log b_i^2 - \sum_{j=1}^{\text N_f^{\vphantom{k}}} f_j^2 \log f_j^2\] \, , \\
&\label{Bk}B_k = \frac{(-1)^k \Gamma\(k - \frac12\)}{\Gamma \(-\frac12\) \Gamma \(k+1\)} \, \zeta\(2 k -1\) \hspace{-2pt} \[\sum_{i=1}^{\text N_b^{\vphantom{k}}} b_i^{2k} - \sum_{j=1}^{\text N_f^{\vphantom{k}}} f_j^{2k} \(2^{2k-1}-1\) \] \, .
\end{align}
\end{subequations}

All in all, $\mathcal G$ takes the asymptotic form
\be
\mathcal G (\varphi, w) \sim 2^{\text N_f} (2\pi)^{\text N_b^{\vphantom{v}}} \, \widetilde C(\varphi, w) \, \mathcal A (\varphi, w) \, e^{ \frac{4 \pi^2 (-\log\hspace{-1pt}w)}{\log^2\hspace{-1.5pt}w + \varphi^2}} \(\log^2\hspace{-1.5pt}w + \varphi^2\)^{4 - \frac{\text N_b^{\vphantom{k}}}{2}} \, , \quad w \to 1^- \, .
\ee
It is thus clear that \eqref{dppwaveZlambda} is the curved-space analogue of the density of states in \eqref{dclflatlambda} equipped with $\mathcal K$ by means of \eqref{Pishift}. Indeed, look at appendix \ref{app:Pi}. Besides the numerical prefactor, the corrections with respect to flat space are encoded in $\tilde C$, $\mathcal A$ and in the power of the $\log$-factor. Anyway, $\tilde C$ and the $\log$-factor do not affect the location and the sharpness of the stationary point ruling the $\varphi$-integral. Moreover, at $\varphi=0$, $\mathcal A$ goes to 1 as $w\to1^-$ at any fixed $\mu$. 
We conclude that the integral over the Lagrange multiplier $\varphi$ results in
\be \label{intKG}
\frac{1}{2\pi} \int_{-\pi}^{+\pi} \hspace{-8pt} d\varphi \,\, \mathcal K \(\varphi,w\) \mathcal G (\varphi,w) \approx \frac{\widetilde C(0, w)}{2^{-\text N_f} (2\pi)^{D-\text N_b^{\vphantom{k}}-10}} \, (-\log w)^{D-\text N_b^{\vphantom{k}}-\frac12} \, e^{-\frac{4\pi^2}{\log w}} \, , \quad w \to 1^- \, .
\ee

Notice that the whole thing admits \eqref{intKPi} as flat space limit, namely
\be \label{flatspacelimit}
\mu , \, \text N_b^{\vphantom{k}}  , \, \text N_f \to 0 \, .
\ee
These limits commute each other. If $\text N_b^{\vphantom{k}}$ ($\text N_f$) vanishes, then there are no massive bosons (fermions) in the model. As a consequence, there are no $b_i \neq 0$ ($f_i \neq 0$). 
In any case, the $\mu\to0$ limit is not enough, since the log-factor in the above formula \eqref{intKG} gets $\mu$-independent correction. However, before applying the identities in \eqref{modidentities}, the latter can be viewed as the $z\to1$ limit of the zero mode contribution in \eqref{dppwavestartcompact}. In turn, this term has no flat space analogue. In other words, it is something that does not reduce to a well-known contribution in flat space in the $\mu\to0$ limit. We believe that this is the reason behind the strangeness of the flat limit defined in \eqref{flatspacelimit}.

So, what we have to compute is
\be \label{dcurvedZlambdaasymp}
d(\mathcal{E}; \mu) \approx \frac{1}{2^{-\text N_f} (2\pi)^{D-\text N_b^{\vphantom{k}}-10}} \int \hspace{-3pt} \rmd\alpha \,\, f(\alpha) \, e^{\Phi(\alpha)} e^{i \Psi(\alpha)}  \, , \quad \mathcal E \to +\infty \, ,
\ee
where we have used the parameterization in \eqref{zpath}
as before and\footnote{Here, we are assuming that $\log \rho <0$. Otherwise, we would have an extra phase factor $e^{\pm i \, (2D-2\text N_b^{\vphantom{k}}-1)\pi/2}$ in $f(\alpha)$. Nevertheless, the case $\log \rho > 0$ would correspond to a minimum of the real part of the integrand, following the path parameterized by \eqref{zpath}; therefore, we can neglect this case.}
\begin{subequations}
\begin{align}
&\Phi(\alpha) = \[\frac{4\pi^2}{(\log \rho)^2 + \alpha^2} + \mathcal{E}\] (- \log \rho) \, ,\\
&\Psi(\alpha) = \[\frac{4\pi^2}{(\log \rho)^2 + \alpha^2} - \mathcal{E} \] \alpha \, ,\\
&f(\alpha) = \widetilde C(0, \rho \, e^{i \alpha}) \[(\log \rho)^2 + \alpha^2\]^{\frac14 \(2D-2\text N_b^{\vphantom{k}}-1\)} e^{\frac{i}{2}\(2D-1-2\text N_b^{\vphantom{k}}\) \arctan\(\frac{\alpha}{\log\rho}\)}\, .
\end{align}
\end{subequations}
Notice that, at least formally, this integral has the very same structure of \eqref{dnflatapprox}, which defines the density of states for an open bosonic string in flat space. What changes is simply the presence of $\mathcal E$ instead of $n$ and the expression for the function $f$. Then, we can recover the discussion about the location and the sharpness of the maximum of $\Phi$ from section \ref{sec:bosopstring}, mapping $n$ into $\mathcal E$.

All in all, the final result is\footnote{Let us stress that the summations run over all the world-sheet modes. Keep in mind that $b_i=0$, $i=N_b^{\vphantom{k}}+1,\ldots,D$, and $f_j=0$, $j=N_f+1,\ldots,D$.}
\vskip -18pt
\begin{subequations}\label{finalresult}
\begin{empheq}[box=\fbox]{align}
&\nonumber \\[-1ex]
&\label{almostfinalresult}
\hspace{5pt}d(\mathcal{E}; \mu) \approx 2^{\text N_f} \, \mathcal C \(\mathcal E, \mu\) \,  \mathcal{E}^{-\frac{D-\text N_b^{\vphantom{k}}+1}{2}} \, e^{4\pi\sqrt{\mathcal{E}}} \, , \quad \mathcal E \to +\infty \, , \hspace{5pt}\\[1ex]
&\label{preC}\hspace{5pt}\mathcal C\(\mathcal E, \mu\) = e^{-2\pi \sqrt{\mathcal E} \, \[1+\sum_{i=1}^8 \(E_0\(\frac{b_i \mu}{\sqrt{\mathcal E}}\)-E_{1/2}\(\frac{f_i \mu}{\sqrt{\mathcal E}}\)\)\]}\, ,\\[-1.5ex]
\nonumber
\end{empheq}
\end{subequations}
where
\be \label{CasimirE}
E_b(m) = \Delta_b(m) + \frac12 \int_{-\infty}^{+\infty} \hspace{-12pt} \rmd k \, \sqrt{\(k+b\)^2+m^2} = \frac12 \sum_{k \in \mathds Z} \sqrt{(k+b)^2+m^2}\, .
\ee
Notice that the above integral is invariant under a shift of $k$ by a constant. Moreover, each $\Delta_b(m)$ is finite and this has been crucial to define the $D$-functions in \eqref{Dfunction}. To the contrary, each $E_b(m)$ is separately divergent, but the summation in \eqref{preC} turns out to be finite and no ad hoc renormalization must be used.\footnote{See \cite{Bigazzi:2003jk} for a very interesting discussion along these lines.} Indeed, from the explicit expression for $\widetilde C$ in \eqref{tildeC} (see also \eqref{C0} and \eqref{subtraction}), we get
\be\label{C}
\mathcal C\(\mathcal E, \mu\) = e^{-\pi\sum_{i=1}^{\text N_b^{\vphantom{k}}} b_i \mu + \frac{2\pi}{\sqrt{\mathcal E}} \log 2 \sum_{i=1}^{\text N_b^{\vphantom{k}}} b_i^2 \mu^2 - 2\pi \sum_{k=2}^{\infty} B_k \, \frac{\mu^{2k}}{\mathcal E^{k-1/2}}}\, .
\ee
Here we kept just the $\mu$-dependent subleading corrections. In fact, due to the (possible) dependence of $\mu$ on $\vec p$, they are the only ones which can survive the large-$\mathcal E$ limit once the integral over the momenta are computed (see next section).

In the flat-space case we have $\mathcal{E}=2n$. Moreover, $\mathcal C$ goes to $1$ as $\mu \to 0$ or $\text N_b^{\vphantom{k}}$, $\text N_f \to 0$. Therefore, the final result \eqref{almostfinalresult} in the limit \eqref{flatspacelimit} smoothly reduces to the outcome we reported in \eqref{dcflattorus} for a closed superstring in flat space times a torus. We can thus rephrase \eqref{almostfinalresult} as
\be \label{finalgenerald}
\boxed{
d(\mathcal{E}; \mu) \sim 2^{\text N_f} \, \mathcal C \(\mathcal E, \mu\) \, \mathcal{E}^{\text N_b^{\vphantom{k}}/2} \, d_{\text{flat$\times$torus}}\(\mathcal{E}\) \, , \quad \mathcal E \to +\infty
}  \, ,
\ee
where $d_{\text{flat$\times$torus}}\(\mathcal E\)$ has been defined in \eqref{dcflattorus}. In this way, the normalization is fixed. The corrections to the constant $d_0$ in the ansatz \eqref{dansatz} and the subleading $\mathcal O \hspace{-2pt} \(1/\sqrt{\mathcal E}\)$ corrections are encoded in the numerical prefactor and in the function $\mathcal C$.

Basically, we have nucleated the corrections to the flat space density of states due to the curvature of the target space. As we anticipated in the introduction, they do not affect the constant $A$ in the ansatz \eqref{dansatz}, but they modify the subleading polynomial behavior encoded in $B$. The latter turns out to be independent of the mass scale $\mu$. Rather, it is fixed by the number of the massive bosonic modes alone. Hence, referring to the flat space limit as defined in \eqref{flatspacelimit} and discussed just below, we have a \emph{smooth} changeover
\be \label{changeover}
B_{\text{flat$\times$torus}} = \frac{D+1}{2} \leftrightarrow \frac{D-\text N_b^{\vphantom{k}}+1}{2} = B \, .
\ee
Possibly, in this sense, that signals the presence of a \emph{continuous} transition between models with different phenomenology in the Hagedorn regime (see next section).

To conclude this section, let us stress that $\mathcal E$ is the eigenvalue of the oscillatory part $\mathcal H$ of the light-cone Hamiltonian defined in \eqref{Hppwave}. Then, the mass-shell condition \eqref{asymmassshell} gives a rule to rephrase \eqref{finalgenerald} as a density of states per unit mass as
\be \label{MtoEpsilon}
\rho(M; \mu) \rmd M = d(\mathcal E; \mu) \rmd \mathcal E \, .
\ee
We get
\be \label{rho}
\rho(M; \mu) \approx 2^{\text N_f} \, 2^{(D-\text N_b)/2} \, \mathcal C \(\alpha' M^2/2,\mu\) \(\frac{1}{\sqrt{\alpha'} M}\)^{\hspace{-4pt}D-\text N_b^{\vphantom{k}}-1} \frac{e^{2\pi\sqrt{2\alpha'} M}}{M} \, , \quad M \to +\infty \, .
\ee
Notice that $2\pi\sqrt{2\alpha'}$ is the leading order value in the $\alpha'$-expansion of the (inverse) Hagedorn temperature for Type II superstring theories in curved space (see \cite{Greene:2002cd, Hyun:2003ks, PandoZayas:2002hh, Grignani:2003cs,Sugawara:2002rs,Brower:2002zx, Bigazzi:2024biz, Harmark:2024ioq, Berkooz:2007fe}). The density of states in \eqref{rho} is such that $\rho(M; \mu) \rmd M$ counts how many single-string states in the single-string spectrum have momentum $\vec p$ and (large) mass between $M$ and $M+\rmd M$. Indeed, let us remember that $\rho$ can depend on the momenta through the mass-scale $\mu$ (hidden in $\mathcal C$).

\section{Strings thermodynamics}
\label{sec:stringstherm}

In general, thermodynamics can arise from either a Lorentzian or an Euclidean perspective. In the former case, the idea is to compute the (canonical) partition function of a thermal system as 
\be \label{canonicalZ}
Z(\beta) = \Tr_{\text{tot}} \, e^{-\beta H} \, ,
\ee
where $\beta$ represents the (inverse) temperature reached at equilibrium, the trace runs over all the possible physical states representing the system itself and $H$ denotes the canonical Hamiltonian whose eigenvalues $E$ are interpreted as energy by means of the temporal derivative of a state.\footnote{Let us stress that such a concept of energy is defined in any (even curved) stationary Lorentzian background. Moreover, it does not depend on the compactification at hand, as it is the case with the definition of mass.} In the latter case, $Z$ depicts a propagation over an imaginary time $i \beta$ from an initial to a final state which are the same due to the presence of the trace. In other words, it corresponds to a vacuum amplitude on a thermal (Euclidean) manifold. 

Of course, these interpretations are just two sides of the same coin and they are both adopted whenever the partition function of the system can be computed. For instance, the reader can take a look at the literature about pp-wave backgrounds \cite{Greene:2002cd, Hyun:2003ks, PandoZayas:2002hh, Grignani:2003cs}. See also a general discussion in \cite{Mertens:2015ola} (in particular, section 2.6).
  
In this section, we aim to discuss the thermodynamic properties of a ``non-interacting'' gas of strings in curved space from a Lorentzian perspective, integrating our novel results with parts of review. As discussed, e.g., in \cite{Sundborg:1984uk, Bowick:1985az, Alvarez:1985fw, Alvarez:1984ee, Alvarez:1986sj}, we can think of a gas of particles at finite temperature sharing the mass spectrum of the string. In other words, if we take a snapshot of the system, each string of the gas will be in a particular state (string mode) which corresponds to a certain state of a particle in the ``analogue'' model. 

Clearly, equilibrium can be reached just thanks to equilibration processes. Strings can split and join through vertex of interactions ruled by the string coupling constant $g_s$. This means that the latter cannot be strictly zero in thermodynamics. The finiteness of $g_s$ makes the number of strings in the gas variable. As a consequence, the grand canonical ensemble would be the most appropriate picture. However, let us stress that string interactions always include gravity and so we must require $g_s$ to be small enough to neglect the backreaction of thermodynamic condensate on the target space. 

As in \cite{Sundborg:1984uk}, we can admit free creation and annihilation of strings, modeling our gas as the extreme equilibrium scenario of thermal radiation with zero chemical potential. In other words, we are assuming to deal with states belonging to a free string spectrum and interactions are included in the model in a crude way just by requiring equilibrium to be established. Accordingly, the results presented for the density of states in the literature hold for a gas of free string. Nevertheless, there is no inconsistency. Indeed, working within the kinetic theory, in \cite{Frey:2023khe} the authors provided a background-independent argument about the equivalence between the equilibrium configurations found out for free strings and the ones computed in string perturbation theory. So the results we referred to apply to the general case of interactions in string perturbation theory. Finally, notice that the grand canonical partition function reduces to the sum of the canonical partition functions \eqref{canonicalZ} for all possible string numbers, since the chemical potential of the strings is assumed to be zero.

In the previous sections, we computed the asymptotic density of states describing the high energy sector of a single free string at fixed center-of-mass momentum. So we already know everything we need to do thermodynamics. Now, it is just a matter of characterizing the ensemble.

The single string explores all the non-compact directions which do not take mass with no obstructions. Indeed, given the embedding map
\be
(\tau,\sigma) \mapsto x^i(\tau,\sigma) \, , \quad i =0,1,...,D-1 \, ,
\ee
such coordinates satisfy massless Klein-Gordon equations on the world-sheet, that is
\be
-\eta^{\alpha\beta} \partial_\alpha \partial_\beta x^\nu = 0 \, , \quad \nu=0,1,...,D-N_b-1 \, ,
\ee
where the Greek indices refer to the world-sheet coordinates $\tau$ and $\sigma$. As a consequence, the general solution to the above equations features a linear term in $\tau$, that is
\be
x^\nu = \alpha' p^\nu \tau + ... \, , \quad \nu=0,1,...,D-N_b-1 \, .
\ee
Lorentz invariance relates the above $p^\nu$ with the $\nu$-th component of the center-of-mass momentum of the string. Then, 
\be \label{masssquaredoperator}
M^2 = - \eta_{\mu\nu} p^\mu p^\nu = \(p^0\)^2 - \vec p^{\,2}
\ee
is interpreted as the squared mass operator of the stringy states, to be fixed by means of the mass-shell condition of the model. The latter can be evaluated on a specific state as in \eqref{asymmassshell}, from which
\be \label{explicitmassshell}
\(p^0\)^2 = \vec p^{\,2} + \frac{2}{\alpha'} \mathcal E \, .
\ee
Let us recall that $\mathcal E$ is the eigenvalue of the operator in \eqref{shiftedH} (see also \eqref{Hppwave}). Therefore, the $D$-dimensional mass $M$ already contains information about the winding numbers and the quantized momenta of the state under examination along the compact directions, besides the occupation numbers of the string oscillators.

On the other hand, the single string experiences a potential in the directions which take mass. Then, it is not completely free to move along them. More explicitly, following the notation of the previous sections, $N_b$ of the embedding directions solve massive Klein-Gordon equations
\be \label{massiveeom}
\(-\eta^{\alpha\beta} \partial_\alpha \partial_\beta + b_i^2 \mu^2\) x^j = 0 \, , \quad j=1,...,N_b \, .
\ee
Clearly, the general solution to the above differential equation does not include a linear term in $\tau$, accordingly to the breaking of Lorentz invariance along these directions.\footnote{The general solution of \eqref{massiveeom} of course displays also a zero-mode part which is $\tau$-dependent. Nevertheless, its contribution is taken into account as zero-mode number operators in the mass-shell condition.} The $j$-th degrees of freedom described by \eqref{massiveeom} is confined in a quadratic potential of characteristic width $(b_j \mu)^{-1}$. As a consequence, in these directions, the string oscillates around a zero-momentum configuration.

In the dual model, $M$ in \eqref{masssquaredoperator} corresponds to the squared mass operator of the particles. It is thus natural to think of the gas of particles as being at equilibrium inside a $(D-\text N_b^{\vphantom{k}}-1)$-dimensional box with reflecting boundary conditions, having length scale $L$ and moving within a reservoir. It is important that the length scale of the box is finite and lower than the Jeans length of the gas. Otherwise, thermal physics would be spoiled by its gravitational collapse. A possible realization of that consists in trading the Minkowskian sector of the target space with an $AdS$ container of radius $L$, as in \cite{Barbon:2001di}. Indeed, the interior part of the $AdS$ space looks flat, while the gas is reflected back over distances comparable with $L$ due the presence of the gravitational potential. 

The trace in \eqref{canonicalZ} runs over all the possible multi-string (or -particle) states of the non-interacting gas. Each of them is given by a collection of single-string (or -particle) states. In turn, a single-string (or -particle) state is fixed by the momentum $\vec p$ of the center-of-mass of the string (or of the particle) and all the other quantum numbers, such as the occupation numbers of the string oscillators, the winding numbers and the quantized momenta in the compact directions. Then, schematically, it can be expanded as the well-known
\be
Z(\beta) = \prod_{b,f} \prod_{k} \frac{1+e^{-\beta E_{f,k}}}{1-e^{-\beta E_{b,k}}} \, ,
\ee
where $b$ and $f$ respectively run over the bosonic and the fermionic single-particle (or -string) states with momentum $k$. The logarithm of the above quantity can be expressed as
\be
\log Z(\beta) = \sum_{r=1}^{+\infty} \frac1r \(Z_{1B}(r \beta) - (-1)^r Z_{1F}(r \beta)\)
\ee
by means of an exact Taylor expansion, where $Z_{1B}$ and $Z_{1F}$ are the bosonic and fermionic single-string partition functions
\be
Z_{1B}(\beta) = \sum_{b} \sum_{k} e^{-\beta E_{b,k}} \, , \quad Z_{1F}(\beta) = \sum_{f} \sum_{k} e^{-\beta E_{f,k}} \, .
\ee
Given a supersymmetric string spectrum, we can assume that
\be
Z_{1B}(\beta) = Z_{1F}(\beta) = Z_1(\beta)/2 \, ,
\ee
from which
\be \label{exactlink}
\log Z(\beta) = \sum_{r=1}^{+\infty} \frac{1}{2 r} \(1 - (-1)^r\) Z_1(r \beta) \, .
\ee

All in all, the partition function of the whole multi-particle (or -string) gas has been expressed in terms of the finite temperature single particle (or -string) partition function
\be \label{beginningZ1}
Z_1(\beta)=\Tr_{\text{single}}\,e^{-\beta \,p^0} \, ,
\ee
where the trace runs over all the single-particle (or -string) states. This provides a perfect playground to explore the consequences of our result on the description of a gas of strings in the Hagedorn regime, taking into account the curvature effects of the target space.

To make contact with the notation of the previous sections, we can use  \eqref{explicitmassshell} and expand the single-string partition function as
\begin{align}
Z_1(\beta)
&=\frac{L^{D-N_b^{\vphantom{k}}-1}}{(2\pi)^{D-N_b^{\vphantom{k}}-1}} \int \hspace{-4pt} \rmd^{D-N_b-1} p \, \, \Tr_{\text{phys}} \, e^{-\beta \,p^0}\\
&=\frac{L^{D-N_b^{\vphantom{k}}-1}}{(2\pi)^{D-N_b^{\vphantom{k}}-1}} \int \hspace{-4pt} \rmd^{D-N_b-1} p \hspace{-2pt} \int \hspace{-4pt} \rmd\mathcal E \, d(\mathcal E; \mu) \,e^{-\beta \, \sqrt{\vec p^{\,2} + \frac{2}{\alpha'} \mathcal E}} \, .
\end{align}
Indeed, $\Tr_{\text{phys}}$ is the trace defined in section \ref{sec:genform} which runs over the physical single-string states at fixed center-of-mass momentum $\vec p$. Let us stress that the trace over the string oscillators (including the zero modes of the massive directions), the winding numbers and the quantized momenta is taken into account by the density of states in the integrand. Finally, using the mass-shell condition \eqref{asymmassshell} and the relation in \eqref{MtoEpsilon}, we get
\be \label{Z1stringframe}
Z_1(\beta)=\frac{L^{D-N_b-1}}{(2\pi)^{D-N_b-1}} \int \hspace{-4pt} \rmd^{D-N_b-1} p \hspace{-2pt} \int \hspace{-4pt} \rmd M \, \rho(M; \mu) \, e^{-\beta \, \sqrt{\vec p^{\,2} + M^2}} \, ,
\ee
where $\rho$ is the density of states per unit mass whose asymptotic form is given in \eqref{rho}. Formally, it looks like the partition function for a gas of free particles in a $(D-\text N_b^{\vphantom{k}})$-dimensional box.

Possibly, we can switch to the more convenient ``light-cone'' frame of reference such that
\be
p^0 = \frac{1}{\sqrt{2}} \(p^+ + \frac{p_T^2+M^2}{2 \, p^+}\) \, , \quad p^\pm= \frac{1}{\sqrt{2}} \(p^0 \pm p^{D-N_b-1}\) \, .
\ee
Here, $\vec p_T$ is the transverse momentum to the light-cone sector. Then, the expressions in \eqref{beginningZ1} and in \eqref{Z1stringframe} can be rephrased as 
\be
Z_1(\beta) = \Tr_{\text{single}} \, e^{-\frac{\beta}{\sqrt{2}}\(p^+ + p^-\)} = \frac{L}{2\pi} \int_0^{+\infty} \hspace{-8pt} \rmd p^+ e^{-\frac{\beta}{\sqrt{2}} p^+} z_1\(\frac{\beta}{2\sqrt{2} \, p^+}\) \, ,
\ee
where the transverse partition function is defined as
\be
z_1(\lambda)=\frac{L^{D-N_b-2}}{(2\pi)^{D-N_b-2}} \int \hspace{-4pt} \rmd^{D-N_b-2} p_T \hspace{-2pt} \int \hspace{-4pt} \rmd M \, \rho(M; \mu) \, e^{-\lambda \(p_T^2 + M^2\)} \, .
\ee

These formulae are totally general and they apply in any regime. Indeed, at least in principle, one can plug into the above expressions for the partition function of the system the general density of states presented in \eqref{dppwavestartcompact}, switching to its version per unit mass by means of \eqref{MtoEpsilon}.
Here, we are interested in the analysis of the Hagedorn regime, where $Z_1$ is expected to break down due to the exponential growth of the density of states at high energies. This is exactly the regime we focused on in the previous sections.
Nevertheless, to perform the computation, we need to know how the dependence of $\mu$ on $\vec p$ is realized in a specific case. Therefore, in the following, we will focus on a huge class of $pp$-wave backgrounds.

\subsection{Type IIA/IIB superstrings on pp-wave geometries}
\label{sec:ppwavethermo}

Here, we focus on solvable plane-wave models with target spaces given by the Penrose limit of global-$AdS_d \times S^n$ spacetimes supported by RR fluxes. The world-sheet description of Type IIA GS superstrings on the non-compact ten-dimensional pp-wave background has been studied in light-cone gauge in \cite{Hyun:2002wu, Hyun:2003ks, Shin:2003ae}. Similar results have been also found in the Type IIB case in \cite{PandoZayas:2002hh, Greene:2002cd, Bigazzi:2003jk}. In other words, the world-sheet spectrum of a closed string probing this kind of backgrounds has already been discussed in the literature. Therefore, we already know everything we need to compute the asymptotic density of states for a large number of cases.

To fix ideas, as $D$-dimensional non-compact sector in the general background \eqref{genback}, we consider the pp-wave geometry given by
\be \label{backexampleppwave}
\rmd s^2_{\text{non-compact}} = - 2 \rmd x^+ \rmd x^- - f^2 \sum_{i=1}^{\text N_b} b_i^2 \, x_i^2 \, (\rmd x^+)^2 +\sum_{j=1}^{D-2} \rmd x_j^2\,.
\ee
In light-cone gauge, the string model is exactly solvable in terms of eight decoupled one-dimensional harmonic oscillators, $\text N_b$ of which have masses $b_i \mu$. Here, $\mu$ is given by
\be \label{mu}
\mu = f \alpha' p^+ \, ,
\ee
where $p^+$ is the light-cone momentum of the string and $f$ is a dimensionful parameter which sets the magnitude of the Ramond-Ramond fluxes supporting the background. We can think of $1/f\sqrt{\alpha'}$ as the curvature length scale of the target space in string units. Finally, by construction, $0 \leq \text N_b \leq D-2$. Anyway, here and in the following, we also consider the $D-\text N_b^{\vphantom{k}}=1$ scenario for completeness. Indeed, in this way we can include the flat space case compactified on a nine-dimensional torus where just one non-compact direction survives and $\text N_b^{\vphantom{k}}=0$. Notice that all the other directions in the compact sector are massless.

For instance, in \cite{Hyun:2002wu, Hyun:2003ks, Shin:2003ae}, the authors provide the world-sheet description of Type IIA Green-Schwarz superstring theory on the ten-dimensional pp-wave background arising as the dimensional reduction of the eleven-dimensional pp-wave geometry given by the Penrose limit of $AdS_4 \times S^7$, with two-form and four-form field strengths. The world-sheet spectrum results in four massive bosons and fermions of mass $\mu/3$ and four massive bosons and fermions of mass $\mu/6$. 

As another example, let us consider the Penrose limit of the Witten background, that is the Type IIA supergravity solution sourced by a stack of $N$ D$4$-branes \cite{Witten:1998zw} dual to the so-called Witten-Yang-Mills (WYM) theory. In \cite{Bigazzi:2004ze}, the authors performed the Penrose limit of this geometry. The resulting world-sheet theory displays three massless bosons, three massive bosons with mass $\mu$ and two massive bosons with mass $\sqrt{3}\mu/2$ (besides eight massive fermions having mass $3 \mu/4$). Here, the field strength of the RR form in $\mu$ is replaced with the mass-scale of the glueballs in the WYM theory.

Finally, in \cite{Russo:2002rq}, the authors focused on Type IIB superstring theories on RR plane wave backgrounds. In particular, they considered the case of plane-wave models with three-form or five-form field strengths. In the former case, the world-sheet spectrum is composed by four massive bosons and four massive fermions with mass equal to $\mu$, in addition to four massless bosons and four massless fermions. In the second case, we have to deal with eight massive bosons and eight massive fermions with again mass equal to $\mu$.

\subsubsection{The single-string density of states per unit energy}
\label{sec:densunitEcomp}

Clearly, within this class of models, the density of states depends just on $p^+$ through $\mu$ in \eqref{mu}. We can thus proceed by massaging the single-string partition function as follows.

First of all, let us introduce the dimensionless variables
\be
b = \beta/f\alpha' \, , \quad m = f \alpha' M \, .
\ee
Performing the integrals over the transverse momenta, we can rephrase $Z_1$ as
\be
Z_1(f\alpha' b) \approx \int \hspace{-4pt} \rmd m \int_0^{+\infty} \hspace{-12pt} \rmd \mu \,\, \mu^{\frac{D-N_b-2}{2}}  \rho\(m/f\alpha'; \mu\)  e^{+b \,S (m, \mu)} \, , \quad S(m, \mu) = -\frac{1}{\sqrt{2}} \(\mu + \frac{m^2}{2\,\mu}\) \, .
\ee

In the Hagedorn regime $\beta \approx \sqrt{\alpha'}$, we have that $b$ is a large parameter in the small curvature limit. Indeed,
\be \label{bHAG}
b \approx 1/f\sqrt{\alpha'} \gg 1 \, .
\ee
Therefore, we can provide an approximated result for the above $\mu$-integral by means of the standard Laplace method. In particular, the function $S$ has a maximum point at
\be \label{maxpoint}
\mu^*=m/\sqrt{2} \, .
\ee
A very important remark is in order. The location of the above maximum point suggests that the final result of the $p^+$-integral is ruled by heavy string states and so the asymptotic expression for $\rho$ in \eqref{rho} can be adopted.\footnote{Indeed, \eqref{maxpoint} implies that the integrand is localized around $p^+ \approx M/\sqrt{2}\Rightarrow p^0 \approx M$, which is the well-known non-relativistic dispersion relation. In \cite{Barbon:2004dd}, the authors concluded that the thermal ensemble beyond string-scale energy densities is dominated by highly excited strings. To get there, they relied on the comparison between the entropies of the two components of the gas, that is the low-energy gravitons and the asymptotic Hagedorn-like tail of the string spectrum. Here, we can see it directly from the thermal partition function.}

All in all, the high-temperature single-string partition function in the small curvature regime can be conveniently expressed as
\be \label{Z1adim}
Z_1(f\alpha' b) \approx \int_{m_0}^{+\infty} \hspace{-12pt} \rmd m \, e^{\sfrac{2\pi\sqrt{2}\,m}{f\sqrt{\alpha'}}} \, m^{-(D-\text N_b^{\vphantom{k}})} \, I(m,b) \, ,
\ee
where
\be \label{I}
I(m,b) = \int_0^{+\infty} \hspace{-12pt} \rmd \mu \,\, \mathcal F(m,\mu) \, e^{+b \, S(m,\mu)} \, , \quad \mathcal F(m,\mu)=\mu^{\frac{D-N_b-2}{2}} \, \mathcal C (m^2/2f^2\alpha', \mu) \, .
\ee
Here, $m_0$ is a threshold such that the approximation in \eqref{rho} is valid.

Then, we just have to compute $I(m,b)$ defined in \eqref{I}. As already pointed out, its structure lends itself to the application of the Laplace method. The latter results in a power series in the small parameter $1/b \ll 1$, that is \cite{Sidorov1985LecturesOT}
\be
I(m,b) \sim \frac{e^{b \, S(m,m/\sqrt{2})}}{\sqrt{b}} \sum_{n=0}^{\infty} \frac{c_n(m)}{b^n} \, , \quad b \to+\infty \, ,
\ee
where\footnote{Superscripts refer to derivatives with respect to the second argument.}
\be
\quad c_n(m) = \frac{\mathcal Q^{(2n)}(m,0)}{(2n)!} \,  \Gamma\(n+\frac12\) \, , \quad \mathcal Q (m,u) = \mathcal F\( m, v(m,u)\) v'(m,u) \, ,
\ee
$v$ being such that
\begin{subequations}
\begin{align}
&S(m,v(m,u))-S(m,m/\sqrt{2}) = - u^2 \, ,\\
&v'(m,0) = \sqrt{\frac{2}{-S''(m,m/\sqrt{2})}} \, .
\end{align}
\end{subequations}
In our case, one can find that
\be
v(m,u) = \frac{m+u^2+u\,\sqrt{2 m + u^2}}{\sqrt{2}} \, .
\ee

Let us proceed step by step. At leading order in the $b$-expansion, we have
\be
c_0(m) = \sqrt{\frac{2\pi}{-S''(m,m/\sqrt{2})}} \, \mathcal F(m,m/\sqrt{2})
\ee
and it is easy to realize that only the first term of $\mathcal C$ in \eqref{C} has to be included at this level of approximation (remember \eqref{bHAG}). We thus get
\be
I(m,b) \sim m^{\frac{D-\text N_b^{\vphantom{k}}-1}{2}} e^{-b \, m } e^{-\pi\sum_{i=1}^{\text N_b^{\vphantom{k}}} b_i \, m/\sqrt{2}} \(1+\mathcal O\(b^{-1}\)\) \, , \quad b\to+\infty \,  .
\ee
Plugging the above result in $Z_1$ and returning to dimensional variables, the result for the single-string partition function is
\be
Z_1(\beta) \approx \int^{+\infty}_{M_0} \hspace{-4pt} \rmd M \, M^{-\frac{D-N_b+1}{2}} \, e^{(\beta_H^{\text{NLO}} - \beta) M} \, ,
\ee
where
\be
\beta_H^{\text{NLO}} = 2\pi\sqrt{2\alpha'} - \frac{\pi}{\sqrt{2}} \, f \alpha' \sum_{i=1}^{\text N_b^{\vphantom{k}}} b_i \, .
\ee
Here, $\beta_H^{\text{NLO}}$ corresponds to the next-to-leading order (NLO) result in the small curvature limit for the (inverse) Hagedorn temperature of Type II superstring theories on RR-supported pp-geometries (e.g., see \cite{Harmark:2024ioq}). 

Notice that $M$ is just a dummy variable and it can be renamed as we want. In particular, we can rephrase the single-string partition function as
\be \label{finalZ1}
Z_1(\beta) \approx \int^{+\infty}_{E_0} \hspace{-8pt} \rmd E \, \, \omega(E) \, e^{- \beta E} \, ,
\ee
where
\be
\omega(E) \approx E^{-\frac{D-N_b+1}{2}} \, e^{\beta_H^{\text{NLO}} E} \, , \quad E \to +\infty \, .
\ee
In this way, $Z_1$ gets the standard Legendre transform structure which comes out from its definition in \eqref{beginningZ1}. Then, $\omega$ can be interpreted as the asymptotic density of states per unit energy $E$. In other words, $\omega(E) \rmd E$ counts how many single-string (or -particle) states have energy between $E$ and $E+\rmd E$.

Going further, let us try to include the first subleading correction of the Laplace method as
\be
I(m,b) \sim \frac{c_0(m)}{\sqrt{b}} \, e^{b \, S\(m, \frac{m}{\sqrt{2}}\)} \[1+\frac{c_1(m)}{c_0(m) b} + \mathcal O \({\frac{1}{b^2}}\)\] \, , \quad b\to\infty \, ,
\ee
or
\be
I(m,b) \approx \frac{c_0(m)}{\sqrt{b}} \, e^{b \, S\(m, \frac{m}{\sqrt{2}}\) + \frac{c_1(m)}{c_0(m) b}} \, , \quad b\to\infty \, .
\ee
It turns out that
\be
\begin{split}
\frac{c_1(m)}{c_0(m)}=&\frac{\pi^2}{4} \({\textstyle{\sum_{i=1}^{\text N_b^{\vphantom{k}}} b_i}} - 4 \,\log 2 \, {\textstyle{\sum_{i=1}^{\text N_b^{\vphantom{k}}} b_i^2}} \, f \sqrt{\alpha'} \)^{\hspace{-2pt}2} \hspace{-2pt} m \, + \\ 
&+ \frac{\pi}{2\sqrt{2}} \[4 \, \log 2 \, {\textstyle{\sum_{i=1}^{\text N_b^{\vphantom{k}}} b_i^2}} \(D-N_b^{\vphantom{k}}+2\)f\sqrt{\alpha'} - \(D - \text N_b^{\vphantom{k}} + 1\)  {\textstyle{\sum_{i=1}^{\text N_b^{\vphantom{k}}} b_i}}\]+ \\[1ex]
&+ \frac{(D-N_b^{\vphantom{k}})^2 - 1}{8\,m} \, .
\end{split}
\ee
Now, we have to plug the above results in  the expression for the single-string partition function according to \eqref{Z1adim}. First of all, let us remember that we are interested in a domain of integration dominated by highly massive states. As a consequence, we can keep track just of the first term of $c_1$. Further, to be consistent with the $f\sqrt{\alpha'}$-expansion of $\beta_H$, this time we have to keep also the second term in $\mathcal C$.  All in all, we get
\be
Z_1(\beta) \approx \int^{+\infty}_{M_0} \hspace{-4pt} \rmd M \, M^{-\frac{D-N_b+1}{2}} \, e^{\(  K(\beta)- \beta\) M} \, ,
\ee
where
\be
K \(\beta\) = 2\pi\sqrt{2\alpha'} - \frac{\pi}{\sqrt{2}} \, f \alpha' \, {\textstyle{\sum_{i=1}^{\text N_b^{\vphantom{k}}} b_i}} + f^2 \alpha'^{3/2} \[ \pi \sqrt{2} \, \log 2 \textstyle \sum_{i=1}^{\text N_b^{\vphantom{k}}} b_i^2 + \frac{\pi^2\sqrt{\alpha'}}{4 \, \beta} \(\textstyle\sum_{i=1}^{\text N_b^{\vphantom{k}}} b_i\)^{\hspace{-2pt}2}\] \, .
\ee
At the Hagedorn point, we have
\be
\beta_H = K(\beta_H) \, ,
\ee
which is solved by
\be \label{betaHNNLO}
\beta_H = 2\pi\sqrt{2\alpha'} - \frac{\pi}{\sqrt{2}} \, f \alpha' \, {\textstyle{\sum_{i=1}^{\text N_b^{\vphantom{k}}} b_i}} + f^2 \alpha'^{3/2} \[\pi\sqrt{2} \,\log 2 \textstyle \sum_{i=1}^{\text N_b^{\vphantom{k}}} b_i^2+ \frac{\pi}{8 \, \sqrt{2}} \(\textstyle\sum_{i=1}^{\text N_b^{\vphantom{k}}} b_i\)^{\hspace{-2pt}2}\] + \mathcal O(f^3 \alpha'^2) \, .
\ee
This matches with the NNLO (inverse) Hagedorn temperature of Type II superstring theories on RR-supported pp-geometries in the small curvature limit (again, see \cite{Harmark:2024ioq}). Notice that the first NNLO term comes from $\mathcal C$ in \eqref{C}, and so it is basically fixed by the  zero point energy of the world-sheet sigma model. On the other hand, the second one comes from the subleading corrections to the result of the $p^+$-integral given by the Laplace method. Notably, the latter has been also computed in \cite{Bigazzi:2024biz} within an effective framework, extending a method which applies to holographic confining backgrounds.

To conclude, with a trivial change of variable in the Hagedorn regime $(\beta-\beta_H)/\beta_H \ll 1$, the single-string partition function can be written exactly as in \eqref{finalZ1}. Now, the single-string density of states per unit energy takes the form
\be\label{densityenergy}
\boxed{
\, \omega(E) \approx \frac{e^{\beta_H E}}{E^{\frac{D-N_b+1}{2}}} \,  \, , \quad E \to +\infty
} \, ,
\ee
$\beta_H$ being the (inverse) Hagedorn temperature presented in \eqref{betaHNNLO}. In general, we envisage that the above expression for the density of states holds regardless the regime we are looking at, interpreting $\beta_H$ as the complete (inverse) Hagedorn temperature of the model.

Apart from the value of the Hagedorn temperature, the single-string partition function \eqref{finalZ1}, equipped with the above density of states, looks like the thermal partition function for a single string embedded in a $(D-N_b)$-dimensional flat space (e.g., see \cite{Alvarez:1985fw, Dienes:1998hx}).
Indeed, formally, our result in \eqref{densityenergy} corresponds to the expression found in \cite{Bowick:1989us} from a world-sheet perspective in a toroidal compactification of flat space, once $D$ is mapped to $D-N_b$.\footnote{In this work, $D$ accounts also for the (non-compact) temporal direction, while in \cite{Bowick:1989us} $D$ represents the number of the spatial non-compact directions alone. For the comparison, the reader should map our $D$ in $D+1$. Let us stress that, in \cite{Bowick:1989us}, the authors worked with $E$ from the beginning and they assumed that it is much greater than the momenta along the compact directions to get the final result. This method has been reviewed also in \cite{Mertens:2015ola}. Here, no such approximations have been taken, since we enclose the data about the compact sector in $\mathcal E$ and so in $M$.} Therefore, it is as if the curvature effects reduce the non-compact directions from $D$ to $D-\text N_b^{\vphantom{k}}$. Accordingly, as we already observed, the string experiences a potential that limits its motion along the directions which take mass (see \eqref{massiveeom}). We can thus define a concept of \emph{effective} non-compact directions as the ones that the string is completely free to explore. This resembles the discussion in \cite{Mitchell:1987hr, Salomonson:1985eq, Lowe:1994nm, Abel:1999rq, Abel:1999dy, Barbon:2004dd, Mertens:2015ola}, where a highly excited free closed string has been modeled as a random walk in the target space given by a spatial toroidal compactification. The conclusion is that the exponent of the energy depends just on the dimension of the volume available for the random walk. Furthermore, this agrees also with the argument in \cite{Horowitz:1997jc}, where the authors derived an expression for the density of states starting from the effective action for the eigenfunction of the string ground state. The latter was supposed to probe a certain numbers ($D - \text N_b^{\vphantom{k}}-1$, in our notation) of “large" spatial directions.

Notably, in curved space, the first corrections to the leading order density of states originate just from the presence of $\text N_b^{\vphantom{k}}$ zero bosonic modes with masses $b_i \mu$, $i=1,...,8$ (see also the expression of $\beta_H$ in \eqref{betaHNNLO}).
This resembles the conclusion drawn in \cite{Bigazzi:2023oqm, Bigazzi:2023hxt,Bigazzi:2024biz}, where the authors computed the first subleading corrections in the $\alpha'$-expansion of the Hagedorn temperature for a large class of confining models. In particular, they noticed how the NLO correction arises entirely from the zero mode part of the massive bosonic world-sheet fields. Notice that it goes along with the discussion in \cite{Greene:2002cd} about the physical interpretation of the result. In few words, string theory is modular invariant and thus the partition function is invariant under transformations which map UV physics in IR physics and vice-versa. This explains why the IR zero modes affect the UV behavior of the asymptotic density of states. Let us stress that the confining models we referred to do not belong to the ones introduced in section \ref{sec:genform}. Anyway, in \cite{Aharony:2003sx}, the authors delved into the thermodynamics of gauge theories on a sphere. In particular, they discussed the phase diagram of large $N$ $SU(N)$ $(\mathcal N = 4)$ Super Yang-Mills theory on $S^3$ at strong 't Hooft coupling (which is ‘‘confining" in the sense explained in \cite{Witten:1998zw}). Intriguingly, they noticed that everything goes as if the target space in the dual stringy description had just one effective non-compact direction. The semiclassical quantization of a closed superstring placed at the center of global-$AdS_5$ produce four massive bosonic modes in the Hagedorn regime. So, this suggests that our formula could have important applications besides solvable string models.

\subsubsection{The canonical and the microcanonical ensemble}
\label{sec:microstuff}
Going further, we have all we need to describe the thermodynamics of a gas of non-interacting strings. In the high temperature regime, we can approximate the multi-string partition function \eqref{exactlink} as\footnote{\label{mbapprox}As it will be clear in the following, the above \eqref{MBZ} realizes in the Maxwell-Boltzmann approximation of the multi-string gas. In \cite{Bowick:1989us}, the authors showed that quantum corrections do not alter the large-energy behavior of the microcanonical density of states we will present here. In fact, it would be modified by a simple multiplicative constant. Therefore, it is totally safe to neglect Bose-Einstein or Fermi-Dirac statistics at high temperature.}
\be \label{MBZ}
Z(\beta) \approx e^{Z_1(\beta)} \, .
\ee
This is clear looking at the single-string partition function in \eqref{finalZ1}, equipped with the density of states \eqref{densityenergy}.

The relevant thermodynamic potentials are
\be \label{canonicalthermo}
F=-\frac1\beta \log Z \, , \quad U = - \frac{\partial \log Z}{\partial\beta} \, , \quad c_V= \beta^2 \frac{\partial^2 \log Z}{\partial \beta^2} \, ,
\ee
respectively the free energy, the internal energy and the specific heat at constant volume of the multi-string gas.
Putting all together, as we approach the Hagedorn temperature from below, the above quantities diverge as\footnote{Notice that for $D-\text N_b^{\vphantom{k}}-2\,n>1$ they are finite and the integral results in $E_0^{1-\mathbb x}/(\mathbb x -1)$, where $\mathbb x=(D-\text N_b^{\vphantom{k}}+1)/2 - n$.}
\be
\int^{+\infty}_{E_0} \hspace{-12pt} \rmd E \,\, \frac{e^{(\beta_H - \beta) E}}{E^{\frac{D-\text N_b+1}{2} - n}} \,  \approx
\begin{cases}
|\log\(\beta-\beta_H\)| \, , & \text{for } D-\text N_b^{\vphantom{k}}-2\,n = 1 \\
\(\beta-\beta_H\)^{\frac{D-\text N_b-1}{2}-n} & \text{for } D-\text N_b^{\vphantom{k}}-2\,n<1
\end{cases} \, , \quad \beta\to\beta_H^+ \, ,
\ee
where $n$ is the order of the derivative with respect to $\beta$ of $\log Z$. Due to the structure of the density of states we found in the previous sections, these behaviors reproduce what has been found in \cite{Dienes:1998hx} for a toroidal compactification of flat space, once $D$ is mapped in $D-\text N_b^{\vphantom{k}}$. Further, as a specific example, let us consider string theory on the Penrose limit of $AdS_5 \times S^5$ \cite{Berenstein:2002jq}. In this case, we have $D=10$ and $\text N_b^{\vphantom{k}}=8$. Therefore, the above results are in perfect agreement with the discussion in \cite{Brower:2002zx}. Indeed, the authors stressed how the free energy matches the one of free strings on an eight-torus:\footnote{Further, they also compare the free energy with the result in \cite{Sundborg:1999ue} for large $N$ $SU(N)$ $(\mathcal N = 4)$ Super Yang Mills on $S^3$. They noticed that the quantities would match if one compactified also the longitudinal direction in the pp-wave geometry. Nevertheless, as we already stressed at the end of the previous section, for a semiclassical string sitting at the center of $AdS_5$ in the Hagedorn regime one would get $\text N_b = 4$, finding agreement with the free energy of large $N$ $SU(N)$ $(\mathcal N=4)$ Super Yang Mills as expected.} in our language, this corresponds to $D-\text N_b^{\vphantom{k}}=2$. 

\begin{table}[t]
\renewcommand{\arraystretch}{1.5}
\centering
 \begin{tabular}{||c c ||} 
 \hline 
 Potential & Diverges as $T_H \to T_H^-$ if  \\ [0.5ex] 
 \hline\hline
 $F$ & $D-\text N_b \leq 1$\\ 
 $U$ & $D-\text N_b \leq 3$ \\
 $c_V$ & $D-\text N_b \leq 5$ \\ [0.5ex] 
 \hline
 \end{tabular}
 \caption{The relevant thermodynamic potentials of the canonical ensemble (listed in \eqref{canonicalthermo}) and the conditions which make them divergent approaching the Hagedorn temperature from below. Case by case, if the inequality is saturated then the divergence is logarithmic. Notice that the above inequalities match the bounds reported in \cite{Dienes:1998hx} for a toroidal compactification of flat space, once $D$ is shifted as $D-\text N_b$.}
 \label{thermotable}
\end{table}

To sum up, we resume in table \ref{thermotable} all the conditions for the divergence of the relevant thermodynamic potentials of the canonical ensemble as we approach the Hagedorn temperature from below.\footnote{As we already stressed in the beginning of section \ref{sec:ppwavethermo}, for the class of backgrounds \eqref{backexampleppwave} we have $D-\text N_b^{\vphantom{k}}\ge2$ by construction. This means that the free energy for a gas of strings probing such backgrounds is never divergent.} In the literature, these divergences have been juxtaposed to a possible ``limiting'' nature of the Hagedorn temperature (e.g., see \cite{Sundborg:1984uk, Alvarez:1985fw, Dienes:1998hx, Greene:2002cd, PandoZayas:2002hh, Brower:2002zx}). Roughly speaking, if the free energy diverges at the Hagedorn point, then an infinite amount of energy would be necessary to raise the temperature of the system above $T_H$. In other words, all the energy we supply to the system is spent on exciting extremely massive string modes. Anyway, the energy fluctuation
\be
\sigma(E) = \frac{\langle E^2 \rangle - \langle E \rangle^2}{\langle E \rangle^2} = \frac{c_V}{\beta^2 U^2}
\ee
diverges as $\beta\to\beta_H^+$ if $D-N_b \leq 5$, which includes all the regimes of table \ref{thermotable}. As a consequence, the energy fluctuation becomes arbitrarily large well before reaching this limiting situation and the canonical ensemble loses reliability: the system gets out of equilibrium and the equivalence between statistical ensembles breaks down. In a sense, under these circumstances,\footnote{Of course, if the energy fluctuations are small at high energies, then there are no troubles and all the ensembles agree to each other.} the Hagedorn regime signals a limit to the existence of matter at equilibrium, even if matter out of equilibrium still (possibly) exists.

Then, a microcanonical description of the system turns out to be more appropriate. Anyway, the canonical formalism does not lose all his utility. Indeed, it can be seen as a tool to derive more easily the density of states per unit energy (as stressed in \cite{Mertens:2015ola}). This is exactly what we have done in \eqref{finalZ1} and \eqref{densityenergy} for the single-string case. Formally, \eqref{finalZ1} can be inverted by means of an inverse Legendre transformation. Then, as observed in \cite{Alvarez:1985fw,Deo:1988jj}, the divergence of the partition function can be avoided by choosing a proper contour of integration in the complex plane. Of course, we can extend this procedure to the multi-string (or -particle) gas. More explicitly,
\be \label{defOmega}
\Omega (E) = \Tr_{\text{tot}} \delta \(E-H\)  = \frac{1}{2\pi i} \int_{\beta_0 - i \, \infty}^{\beta_0 + i \,\infty} \hspace{-4pt} \rmd \beta \, \, Z (\beta) \, e^{+\beta E} \, , \quad \beta_0 > \beta_H \, ,
\ee
is the microcanonical density of states per unit energy such that $\Omega(E) \rmd E$ counts how many multi-string (or -particle) states have energy between $E$ and $E+\rmd E$, where $Z$ is defined in \eqref{canonicalZ}.

In the high temperature regime, we can plug into the definition of the microcanonical density of states in \eqref{defOmega} the approximated version of $Z$ in \eqref{MBZ}. Once the exponential of the single-string partition function has been expanded, a change of variable makes it easy to reconstruct the integral representation of the Dirac delta. The final result is\footnote{Notice that each $E_i$ cannot exceed $E$, otherwise the Dirac delta would make that contribution vanish. This is the reason why the upper extremes of integration have been all reduced from $+\infty$ to E.}
\be
\Omega(E) = \sum_{n=0}^{\infty} \Omega_n(E) \, , \quad \Omega_n(E) = \frac{1}{n!} \prod_{j=1}^{n} \int_{E_0}^E \hspace{-6pt} \rmd E_j \, \omega(E_j) \, \delta\(E-\sum_{k=1}^n E_k\) \, .
\ee
This is the usual Maxwell-Boltzmann expression for a classical system (as we have anticipated in footnote \ref{mbapprox}). Indeed notice that $\Omega_n(E)$ represents the degeneracy of states featuring $n$ particles (or strings) and total energy $E$. The knowledge of $\omega$ in \eqref{densityenergy} allows us to perform its computation explicitly, at least in the high-energy regime. Notice that just one Dirac delta appears. Moreover, if $E$ equalized between two or more (let us say $m$) $E_j$, then morally the integrand would scale as $e^{\beta_H E}/E^{m(D-N_b+1)/2}$. In the large-$E$ limit, these configurations are clearly suppressed with respect to one of the $n$ which displays just one $E_i \approx E$ for a certain $i$. It thus follows that
\be\label{Omegan}
\Omega_n(E) \approx \frac{1}{(n-1)!} \frac{e^{\beta_H E}}{E^{(D-N_b+1)/2}} \(\int_{E_0}^E \frac{\rmd E'}{E'^{(D-N_b+1)/2}}\)^{\hspace{-4pt}n-1} \, , \quad E \to +\infty \, ,
\ee
and so
\be\label{Omegancases}
\Omega_n(E) \approx
\begin{cases}
\displaystyle \frac{1}{(n-1)!} \frac{e^{\beta_H E}}{E^{(D-N_b+1)/2}} \(\frac{2}{(D-N_b-1) E_0^{(D-N_b-1)/2}}\)^{\hspace{-4pt}n-1} \, , \quad &\text{for } D-N_b>1 \, , \\
\displaystyle \frac{1}{(n-1)!} \frac{e^{\beta_H E}}{E^{(D-N_b+1)/2}} \(\log(E/E_0)\)^{n-1} \, , \quad &\text{for } D-N_b=1 \, .
\end{cases}
\ee
Summing over all the possible $n$, we get\footnote{The $n=0$ contribution gives a $\delta(E)$, which can be neglected since we are assuming that $E$ is significantly different from zero.}
\begin{subequations}\label{finalOmega}
\begin{align}
&\label{Omegapeak}\Omega(E) \approx \frac{e^{\beta_H E}}{E^{\frac{D-N_b+1}{2}}} \,  e^{\frac{2}{(D-N_b-1) E_0^{(D-N_b-1)/2}}}  \, ,  &&\hspace{-12pt}D-N_b>1 \, ,  \quad E \to +\infty \, ,\\
&\Omega(E) \approx \frac{e^{\beta_H E}}{E_0} \, ,  &&\hspace{-12pt}D-N_b=1 \, ,  \quad E \to +\infty \, .
\end{align}
\end{subequations}

Again, our result for $\Omega$ corresponds to the well-known outcome in flat space, once $D$ is mapped to $D-N_b$ \cite{Bowick:1985az, Brandenberger:1988aj, Deo:1988jj, Deo:1989bv, Deo:1991mp, Deo:1991af, Mertens:2015ola}. With this result at hand, in principle we could compute the energy distribution function $\mathcal D(\epsilon, E)$ in the microcanonical ensemble, such that $ \mathcal D(\epsilon, E) \rmd \epsilon$ gives the average number of strings having energy between $\epsilon$ and $\epsilon+\rmd \epsilon$ in a system of total energy $E$. Anyway, just looking at $\Omega$ we already have what we need here. Indeed, apart from a constant overall factor, the multi-string density of states in \eqref{Omegapeak} has the same formal expression of the single-state density of states in \eqref{densityenergy}. This means that the energy of the multi-string gas is basically gathered in a single long string for $D-N_b>1$.\footnote{Notice that a highly exited string is represented by a very massive state. Then, its energy should be well described by classical arguments as $E\approx l/\alpha'$, $l$ being the length of the string. Therefore, at least for $D-N_b>1$, highly excited strings has the tendency to join together becoming longer and longer.} On the other hand, the two expressions differ a bit for $D-N_b=1$ and we cannot come to the same conclusion. Morally, from \eqref{Omegancases}, we can argue that the contribution of the other $n-1$ strings is $E$-dependent and logarithmically large at high energy. So, even if \eqref{Omegan} derives from $E\approx E_i$ for a given $i$, the summation over all the possible $n$ makes more uniform configurations come into play.\footnote{For $N_b=0$, the reader can find a detailed analysis of this scenario in \cite{Brandenberger:1988aj, Deo:1989bv}. In particular, \cite{Brandenberger:1988aj} is totally devoted to the $D=1$ case. On the other hand, in \cite{Deo:1989bv} the authors delved into the general case in full-details. Let us stress that here we neglected subleading corrections to \eqref{densityenergy}, as in \cite{Bowick:1985az, Mitchell:1987hr, Salomonson:1985eq, Lowe:1994nm, Abel:1999rq, Abel:1999dy, Barbon:2004dd, Mertens:2015ola}. In \cite{Deo:1989bv}, the authors showed that in lower dimensions (namely, $D=2,3$) such corrections can affect the final result for $\Omega$ which no longer turns out to be dominated by a single string. Anyway, these deviations strongly depend on the prescription adopted for the thermodynamic limit and they vanish in the case of arbitrarily large (strictly infinite) energy density. Therefore, here we do not consider these effects, but the reader should be aware of them. For completeness, the reader should also be aware of \cite{Deo:1991mp}. There, the authors showed that if the target space is compact with radius $R$ but large ($R \gg \sqrt{\alpha'}$), then at finite energy density other $R$-dependent corrections arise if $E \ll R^2 \alpha'^{-3/2}$. Again, all these effects can be omitted if we think about arbitrarily high energies.}

With this information at hand, we can discuss the phase diagram of the model around the Hagedorn temperature. Indeed, the Boltzmann entropy
\be
S=\log\(\Omega(E) \Delta E \) \, ,
\ee
$\Delta E$ being a constant with the dimension of energy, induces a formal definition of temperature as\footnote{Usually, this is the condition which realizes the equivalence of the canonical and the microcanonical ensembles in the thermodynamic limit. Indeed, they are put in contact through the saddle point approximation of $Z(\beta)=\int \hspace{-4pt} \rmd E \, \Omega(E) \, e^{-\beta E}$. Formula \eqref{temperature} defines the location of the saddle point. Of course, this makes sense if the energy fluctuations in the canonical ensemble are small.}
\be \label{temperature}
\frac1T = \frac{\partial S}{\partial E} \, .
\ee

\begin{figure}[t]
	\begin{center}
		\scalebox{0.275}{
		\includegraphics{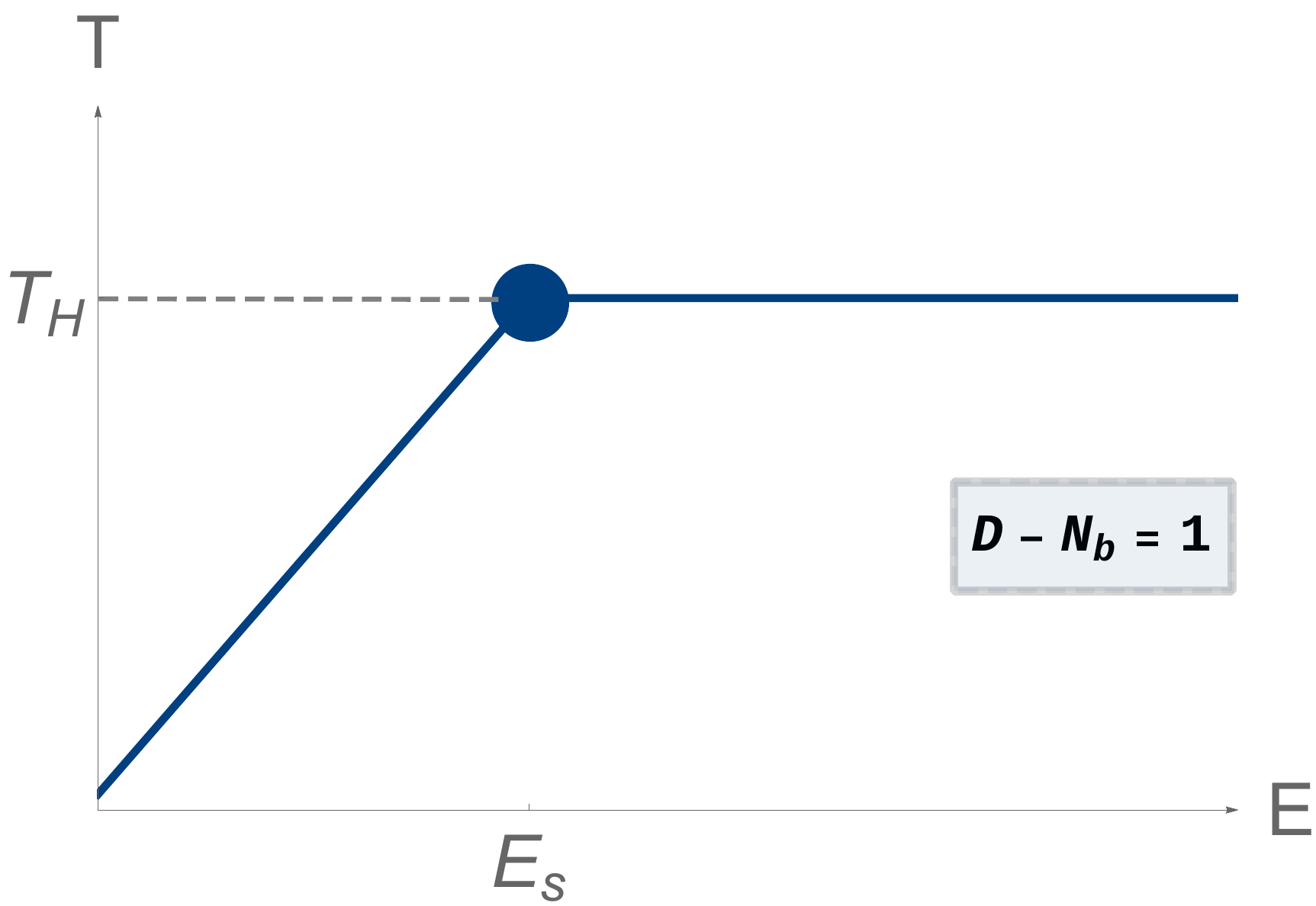}
		}
		\scalebox{0.275}{
		\includegraphics{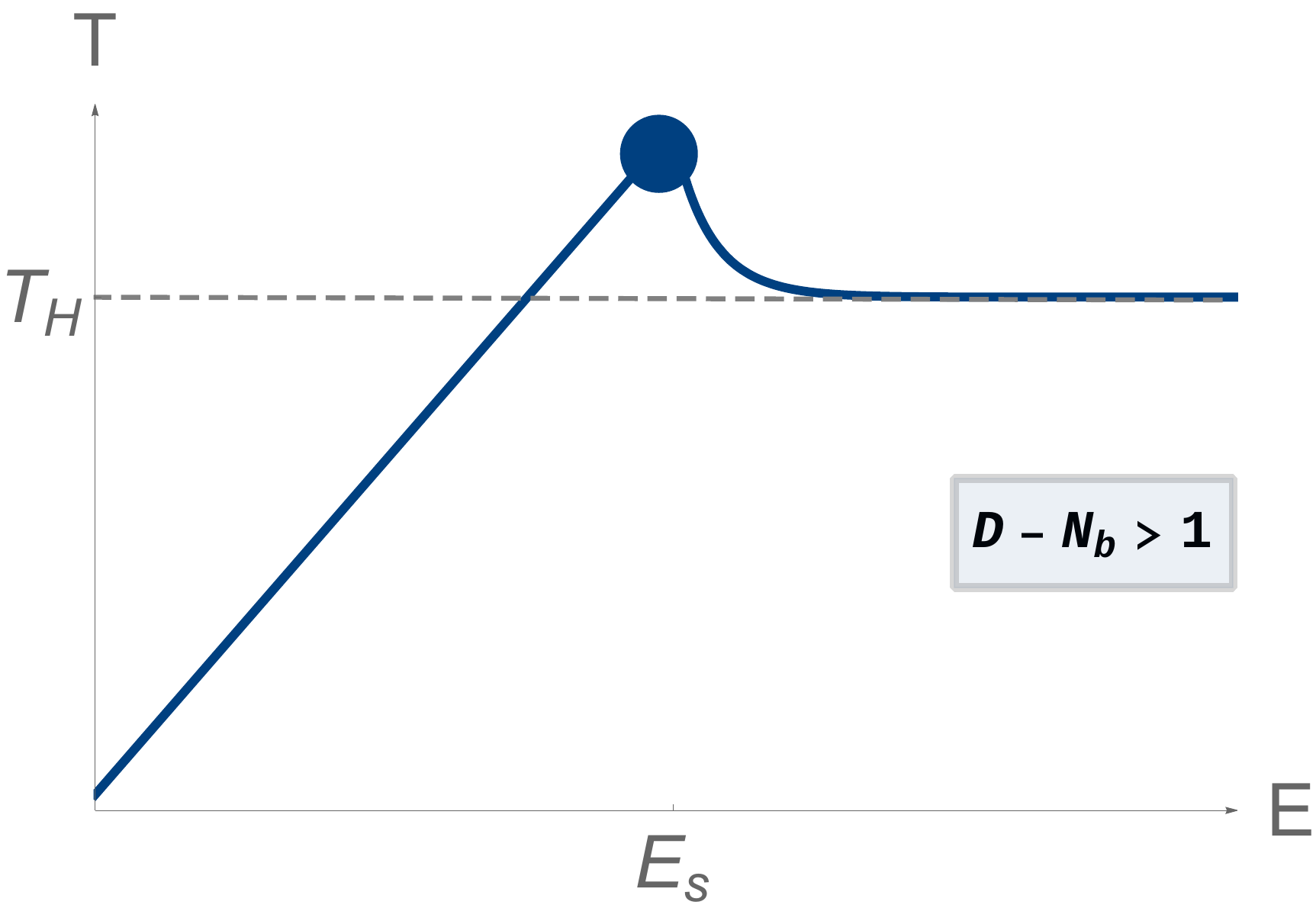}
		}
	\end{center}
	\caption{Plot of the microcanonical temperature $T$ of the multi-string gas defined in \eqref{temperature} against the total energy of the ensemble $E$, around the Hagedorn point. At low energies, the gas is dominated by massless modes, while at high energies the highly massive string states come into play. The transition happens somewhere around the string scale $E_s \sim 1/\sqrt{\alpha'}$. Notice that, if $D-\text N_b>1$, $T$ slightly overshoots the Hagedorn temperature $T_H$, resulting in a locally unstable phase dominated by a single long string. On the other hand, within the current approximations, if $D-\text N_b=1$ the diagram features a phase with strictly infinite specific heat and the total energy is more equally distributed among the strings of the gas.}
	\label{fig:TcompactVsunompact}
\end{figure}
So far, we focused just on the highly excited string modes which are responsible for the exponential growth of the asymptotic density of states reported in \eqref{finalOmega}. Anyway, the whole spectrum features also a component of massless modes having entropy of order $E^{(D-1)/D}$.\footnote{The entropy and the energy of a gas of massless particles at a given temperature $T$ in a $(D-1)$-dimensional box of volume $V$ are extensive. Therefore, we must have $S \sim V T^{D-1}$ and $E \sim V T^D$. So the dependence of $S$ on $E$ follows by a simple dimensional analysis.} Then, the dependence of the temperature on the energy can be sketched out as in figure \ref{fig:TcompactVsunompact}, by choosing at each value of the total energy the most entropic phase between the two just described. 

The massless modes dominate the ensemble at low energies, while the exponentially-growing massive states are entropically favored beyond the string scale. Notice that the diagram features an horizontal plateau at the Hagedorn temperature. As a consequence, all the energy injected into the system is spent on exciting increasingly massive states, instead of rising the temperature of the gas. Thus, $T$ is basically kept fixed and 
bounded from above. Moreover, if at least one of the spatial directions is \emph{effectively} non-compact,\footnote{Let us remember that the number of effective non-compact (temporal+spatial) directions is $D-N_b$.} then the Hagedorn plateau corresponds to a locally unstable phase dominated by a single long string (possibly out of equilibrium, see before). In this sense, we may interpret the Hagedorn temperature as the maximum allowed physical temperature (as proposed in \cite{Brandenberger:1988aj}).

What we can do with our results is to better resolve the Hagedorn plateau. If the target space is such that $D-N_b=1$ than the branch is truly horizontal and describes a phase of uniformly energetic strings with infinite specific heat $c_V = (\partial T/ \partial E)^{-1}$. On the other hand, if $D-N_b>1$, the temperature slightly overshoots the Hagedorn value to which it tends from above as the energy increases. Therefore, we can think about this phase as a single folded long string with negative specific heat. This corresponds to one of the two main qualitative scenarios reported in \cite{Barbon:2004dd} (see the locally unstable branch in figure 7).

\begin{figure}[t]
	\begin{center}
		\scalebox{0.24}{
		\hspace{-38pt}\includegraphics{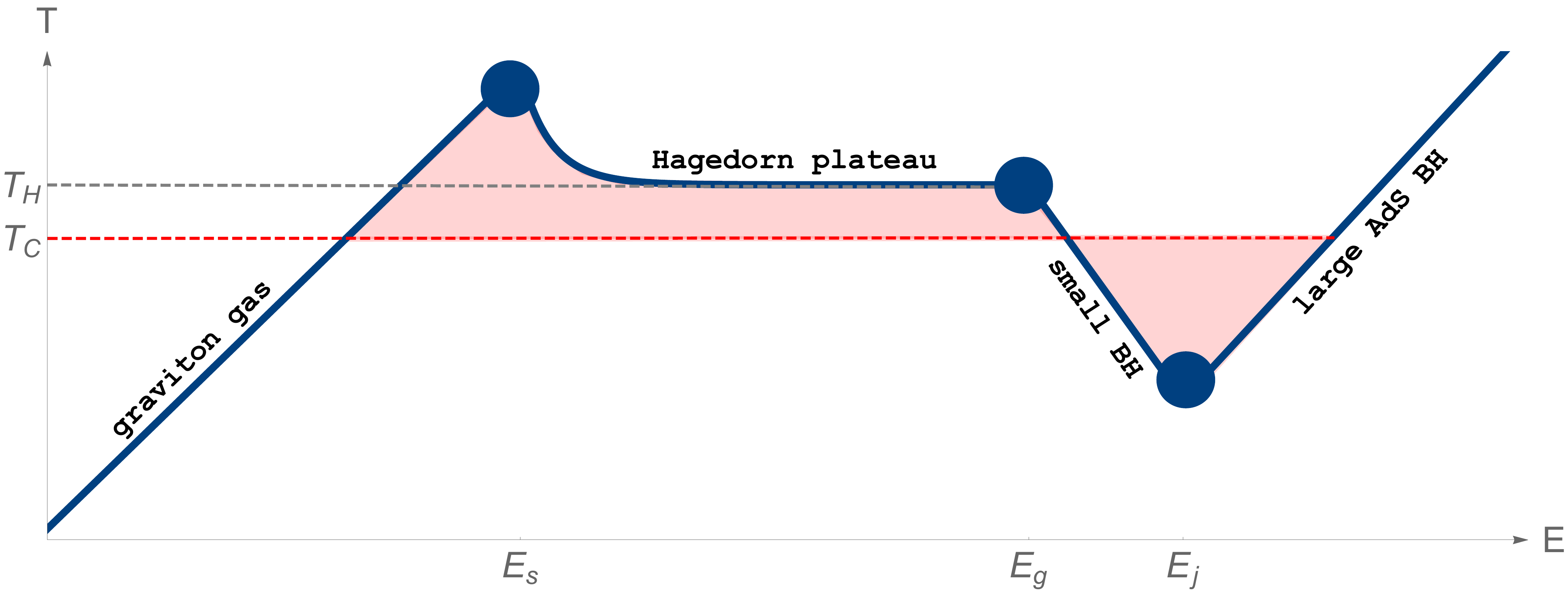}
		}
		\caption{Temperature versus energy diagram of the gas of strings which are assumed to collapse into black holes at high energies, once the Jeans instability has been regularized through an $AdS$ container. The first (from the left) two phases already introduced in figure \ref{fig:TcompactVsunompact} are followed by an unstable phase of Schwarzschild black holes and a stable phase of large black holes in $AdS$. Here, $T_c$ represents the critical temperature of a first-order phase transition among the graviton gas and the $AdS$ black holes. The Maxwell construction in red makes it clear that the Hagedorn plateau is a superheated phase missed by the canonical ensemble.
		}
		\label{fig:phasediagram}
	\end{center}
\end{figure}
In any case, the temperature is approximately independent of the energy and so the resulting specific heat is very large, signaling a would-be never-ending plateau. Nevertheless, relying on the principle of ``asymptotic darkness” \cite{tHooft:1984kcu, Banks:2003vp}, we can expect that, at extremely large energies, black holes rule the phase diagram of models which include gravity. This scenario has already been discussed in the literature \cite{Barbon:2001di, Barbon:2004dd, Abel:1999dy, Aharony:2003sx, Mertens:2015ola}. Since thermodynamics seems to work here as in a $(D-\text N_b)$-dimensional flat space, let us review well-known results extending them to the current framework in a naive way. Notice that the entropy of Schwarzschild black holes in $D-\text N_b-1$ dimensions scales as
\be
S_{SBH} \sim E \, (g_s^2 E)^{\frac{1}{D-\text N_b-3}} \, .
\ee
Then, the Hagedorn branch $S\sim\beta_H E$ is expected to end at $\sqrt{\alpha'} E_g \sim 1/g_s^2$. What follows should be a phase of small black holes with negative specific heat for $D-\text N_b>3$.\footnote{For $D-\text N_b=2,3$, the “asymptotic darkness” gets more involved as localized energy sources have an impact on the asymptotic conditions for the vacuum. Here, we do not consider these cases, since a detailed description of the phase diagram beyond the Hagedorn plateau exceeds the purposes of this work.} The latter is bound to end whenever such black holes fill all the available space in the box on scales of the Jeans length. Going further, we can think about the box as an $AdS$ container,\footnote{Here, we are focusing on $1/N \ll g_s \ll 1/N^{1/5}$, $N$ being the number of colors in the dual Yang Mills theory by means of the $AdS$/CFT correspondence \cite{Barbon:2001di, Mertens:2015ola}. Indeed, given the holographic dictionary, one can show that $g_s N^{1/5} \gtrsim 1$ would imply the $AdS$ radius to exceed the Jeans length. In such a case the Hagedorn plateau would not be there, since the gravitational collapse would occur too soon. On the other hand, if $g_s N \lesssim 1$, then $L \lesssim \sqrt{\alpha'}$ and a dual description in terms of Yang Mills variables would be more appropriate~\cite{Aharony:2003sx}. Notice that $g_s$ is small but not strictly zero, in agreement with the discussion at the very beginning of the current section.} which thus plays the role of an IR regulator of the Jeans instability. In this picture, a (stable) phase of large $AdS$ black hole is envisaged to take over from the Jeans scale $\sqrt{\alpha'} E_j \sim (L/\sqrt{\alpha'})^{D-\text N_b-3}(1/g_s^2)$ on. In figure \ref{fig:phasediagram}, we report the complete microcanonical temperature as a function of the energy.

A Maxwell construction in the diagram of figure \ref{fig:phasediagram} gives a first order phase transition at a critical temperature which is lower than the Hagedorn value. Then, very large black holes are nucleated directly out of the low energy phase of massless modes. As a consequence, the Hagedorn plateau must be interpreted as only accessible to a superheated phase which is missed in the canonical picture, as well as the unstable phase of small black holes. This phenomenon has been introduced under the name of ``Hagedorn censorship'' in \cite{Barbon:1998ix, Barbon:1998cr, Barbon:2001di} and mitigates the (possible) tension among the canonical and the microcanonical ensembles, as we are describing a transition among phase at equilibrium which shadows the problematic region of the diagram in figure \ref{fig:phasediagram}.

\section{Conclusions}
\label{sec:conclusions}

In this paper, we computed the asymptotic density of states for a class of solvable superstring models on curved backgrounds without running dilaton. In particular, we provided the corrections with respect to the well-know result in flat space coming from the curvature of the target space.

At fixed center-of-mass momentum $\vec p$, the final result of our computation is
\be
d(\mathcal{E}; \mu) \approx 2^{\text N_f} \, \mathcal C \(\mathcal E, \mu\) \,  \mathcal{E}^{-\frac{D-\text N_b^{\vphantom{k}}+1}{2}} \, e^{4\pi\sqrt{\mathcal{E}}} \, , \quad \mathcal E \to +\infty \, ,
\ee
where
\be
\mathcal C\(\mathcal E, \mu\) = e^{-2\pi \sqrt{\mathcal E} \, \[1+\sum_{i=1}^8 \(E_0\(\frac{b_i \mu}{\sqrt{\mathcal E}}\)-E_{1/2}\(\frac{f_i \mu}{\sqrt{\mathcal E}}\)\)\]}\, ,
\ee
as already reported in \eqref{finalresult}. For the reader convenience, remember that $\mathcal E$ is the eigenvalue of the oscillatory part of the light-cone Hamiltonian introduced in \eqref{Hppwave}. Moreover, $D$ represents the number of non-compact directions in the background, while $\text N_b^{\vphantom{k}}$ $\(\text N_f\)$ counts the number of massive bosons (fermions) in the world-sheet theory. Further, $b_i$ ($f_i$) is the mass of the $i$-th bosonic (fermionic) mode in units of the dimensionless mass-scale $\mu$, which (in general) can depend on $\vec p$. Finally, $E_0$ and $E_{1/2}$ are the zero-point energies defined in \eqref{CasimirE} (see also the expansions in \eqref{Casimirexpansion}). Notice that, looking at the explicit expression for $\mathcal C$ in \eqref{C}, the above result perfectly matches the expectation we presented in the ansatz \eqref{dansatz} and extends the proposal in \cite{Russo:2002rq} beyond the leading exponential behavior.\footnote{In principle, other $\mathcal O\(1/\sqrt{\mathcal E}\)$-corrections could arise from the computations. Here, we focused just on the $\mu$-dependent ones: as we already pointed out in the main body, these are the only which can survive the large-$\mathcal E$ limit thanks to the (possible) dependence of $\mu$ on $\vec p$. See section \ref{sec:densunitEcomp} for details.}

The computation of the partition function of the model requires to trace over all the possible momenta of the single string. Therefore, the dependence of $\mu$ on $\vec p$ turns out to be fundamental. For a string on a pp-wave geometry, it is well-known that $\mu$ is proportional to the (inverse) curvature scale length $f$ of the background and to the light-cone momentum $p^+$ of the string itself. At least in the small curvature regime, this has been enough to extract the asymptotic density of states per unit energy as
\be
\omega(E) \approx \frac{e^{\beta_H E}}{E^{\frac{D-N_b+1}{2}}} \,  \, , \quad E \to +\infty \, ,
\ee
where
\be
\beta_H = 2\pi\sqrt{2\alpha'} - \frac{\pi}{\sqrt{2}} \, f \alpha' \, {\textstyle{\sum_{i=1}^{\text N_b^{\vphantom{k}}} b_i}} + f^2 \alpha'^{3/2} \[\pi\sqrt{2} \, \log 2 \textstyle \sum_{i=1}^{\text N_b^{\vphantom{k}}} b_i^2 + \frac{\pi}{8 \, \sqrt{2}} \(\textstyle\sum_{i=1}^{\text N_b^{\vphantom{k}}} b_i\)^{\hspace{-2pt}2}\] + \mathcal O(f^3 \alpha'^2)\, .
\ee
Here, $\beta_H$ represents the next-to-next-to-leading order (NNLO) value in the small curvature limit of the (inverse) Hagedorn temperature for Type II superstring theories on the backgrounds introduced in section \ref{sec:ppwavethermo}. Notice that the LO contribution is fixed by the leading exponential factor in $d$, already envisaged in \cite{Russo:2002rq}. On the other hand, the subleading corrections originate from the $\mu$-dependent contributions encoded in the function $\mathcal C$ above.

Once the proper value of $\beta_H$ is adopted (i.e. including the curvature corrections), everything goes as if the string probed a $(D-\text N_b)$-dimensional flat spacetime. This follows from the $\text N_b$-dependent correction in the polynomial behavior of the final result for the density of states.\footnote{Again, this correction was missing in \cite{Russo:2002rq}.} The reader can find a detailed analysis of the thermodynamic consequences of this interpretation in section \ref{sec:microstuff}.

To conclude, our proposal aims at shedding some light on a compelling open problem of string theory by means of a first principle computation. This is interesting in itself, but equally interesting applications lie ahead. Indeed, notice that the computation relies just on the quadratic structure of the string canonical Hamiltonian and the absence of a running dilaton. Therefore, it can be applied to any scenario featuring these properties. For instance, we can consider a generic non-linear world-sheet sigma model expanded up to second order in the small fluctuations around a reference classical configuration, such that the dilaton takes a trivial profile. Of course, the resulting spectrum hosts states coming from just a particular sector of the whole theory. Nevertheless, it is the best we can do with this formalism. We defer applications along these lines in a future work.

\vskip 15pt \centerline{\bf Acknowledgments} \vskip 10pt 

\noindent 
We are indebted to Francesco Bigazzi, Matteo Ciardi, Filippo Colomo, Aldo Cotrone, Troels Harmark, Wolfgang Mück, Andrea Olzi, Jorge Russo, Arkady Tseytlin and Riccardo Villa for comments and very helpful discussions. In particular, we really want to thank Troels Harmark for inspiring this revised version with his remarks.

\appendix

\section{Jacobi Theta Functions}
\label{app:theta}

In this appendix, for the reader convenience, we resume all the properties of the Jacobi Theta Functions which have been used throughout this paper. These four particular functions are defined as Fourier series as \cite{NIST:DLMF}
\begin{align}
&\theta_1(z,q) = 2 \sum_{n=0}^\infty (-1)^n q^{(n + 1/2)^2} \sin((2n+1)z) \, , \\
&\theta_2(z,q) = 2 \sum_{n=0}^\infty q^{(n+1/2)^2} \cos((2n+1)z) \, , \\
&\theta_3(z,q)=1+2\sum_{n=1}^\infty q^{n^2} \cos{2nz} \, , \\
&\theta_4(z,q)= 1+2\sum_{n=1}^\infty (-1)^n q^{n^2} \cos{2nz} \, .
\end{align}
Denoting with the apostrophe the derivative with respect to $z$, they all fulfill the mirror symmetry~\cite{Wolfram}
\be
\theta_j(z^*,q^*)=\theta_j(z,q)^* \, , \quad \theta'_j(z^*,q^*)=\theta'_j(z,q)^* \, , \quad j=1,2,3,4 \, .
\ee
Moreover, at $z=0$ it follows that \cite{NIST:DLMF}
\begin{subequations} \label{thetazzero}
\begin{align}
&\theta'_1(0,q) = 2 \, q^{1/4} \prod_{n=1}^\infty (1-q^{2n})^3 \, , \\
&\theta_2(0,q) = 2 \, q^{1/4} \prod_{n=1}^\infty (1-q^{2n}) (1+q^{2n})^2 \, , \\
&\theta_3(0,q)=\prod_{n=1}^\infty (1-q^{2n}) (1+q^{2n-1})^2 \, , \\
&\theta_4(0,q)= \prod_{n=1}^\infty (1-q^{2n}) (1-q^{2n-1})^2 \, .
\end{align}
\end{subequations}
Remarkably, they satisfy
\be
\theta'_1(0,q) = \theta_2(0,q) \, \theta_3(0,q) \, \theta_4(0,q) \, .
\ee
In general, the expressions in \eqref{thetazzero} can be rephrased in terms of the Dedekind eta function. In particular what we need is \cite{Wolfram}
\begin{subequations}\label{asymtheta}
\begin{align}
&\theta_2(0,q) = \frac{2}{\eta\(-\frac{i}{\pi}\log q\)}\eta{\(-\frac{2\,i}{\pi} \log q\)^{\hspace{-3pt}2}} \sim \(\frac{\pi}{-\log q}\)^{\hspace{-2pt}\frac{1}{2}} \, , \quad q\to1^- \, , \\
&\label{asymtheta3}\theta_3(0,q) = \frac{\eta{\(-\frac{i}{\pi} \log q\)^5}}{\eta{\(-\frac{2\,i}{\pi} \log q\)^2} \eta{\(-\frac{i}{2\pi} \log q\)^2}} \sim \(\frac{\pi}{-\log q}\)^{\hspace{-2pt}\frac{1}{2}} \, , \quad q\to1^- \, , \\
&\label{asymtheta4}\theta_4(0,q) = \frac{1}{\eta\(-\frac{i}{\pi}\log q\)}\eta\(-\frac{i}{2\pi} \log q\)^{\hspace{-3pt}2} \sim 2 \(\frac{\pi}{-\log q}\)^{\hspace{-2pt}\frac12} e^{\frac{\pi^2}{4\,\log q}} \, , \quad q\to1^- \, .
\end{align}
\end{subequations}

\section{The generating function in flat space}
\label{app:Pi}

Here, we want to give an explicit expression for the function $\Pi$ appearing in \eqref{dclflatlambda}. Moreover, we will prove the result given in \eqref{ansatz}. After that, we will also discuss how such result is affected by a possible toroidal compactification.

To begin with,
\begin{align}
\prod_{k=1}^\infty \, \left | \frac{1 + q^k}{1-q^k } \right |^{16} 
&= \prod_{n=1}^\infty \, \left | \frac{1 + q^{2n}}{1-q^{2n}} \right |^{16} \prod_{m=1}^\infty \, \left | \frac{1 + q^{2m-1}}{1-q^{2m-1}} \right |^{16}\nb \\
&= \prod_{n=1}^\infty \left ( \frac{1 + q^{2n}}{1-q^{2n}} \right )^{\hspace{-4pt}8} \( \frac{1 + (q^*)^{2n}}{1-(q^*)^{2n}} \)^{\hspace{-4pt}8} \prod_{m=1}^\infty \( \frac{1 + q^{2m-1}}{1-q^{2m-1}} \)^{\hspace{-4pt}8} \( \frac{1 + (q^*)^{2m-1}}{1-(q^*)^{2m-1}} \)^{\hspace{-4pt}8} \, ,
\end{align}
where we used the well-known property of the complex conjugation
\be
(q^n)^* = (q^*)^n \, , \quad \forall \, q \in \mathds C \, , \quad \forall \, n \in \mathds Z \, .
\ee
Therefore, given the features of the Jacobi Theta Functions resumed in appendix \ref{app:theta}, we conclude that
\be
\Pi(\varphi,w) = \left | \frac{\theta_2\(0, w \, e^{i\,\varphi}\) \theta_3\(0, w \, e^{i\,\varphi}\)}{\theta'_1\(0, w \, e^{i\,\varphi}\) \theta_4\(0, w \, e^{i\,\varphi}\)} \right |^8 = \frac{1}{\left | \theta_4\(0, w \, e^{i \, \varphi}\) \right |^{16}} \, .
\ee

With this expression at hand, 
we can check formula \eqref{ansatz}. Let us take
\be
w=e^{-2\pi/\sqrt{\mathcal{E}}} \to 1^- \, , \quad \mathcal{E} \to +\infty \, .
\ee
A well-known formula (e.g., see \cite{Sundborg:1984uk}; cf.~\eqref{asymtheta4}) states that
\be \label{thetasundborg}
\theta_4(0,w \, e^{i \varphi})\sim 2 \(-\frac{\log w + i \varphi}{\pi}\)^{\hspace{-2pt}-1/2} e^{\pi^2/ 4(\log w + i \varphi)} \, , \quad (w \to 1^-  , \, \varphi \to 0) \, .
\ee

All in all, we have
\be
\frac{1}{2\pi} \int_{-\pi}^{+\pi} \hspace{-4pt} d\varphi \,\, \Pi(\varphi,w) \sim \frac{1}{2^{17}\pi^9} \int_{-\pi}^{+\pi} \hspace{-4pt} d\varphi \, \, Q(\varphi,w) \, e^{P(\varphi, w)} \, , \quad w\to1^- \, ,
\ee
where
\be
P(\varphi, w) = -4\pi^2 \frac{\log w}{(\log w)^2 + \varphi^2} 
\, , \quad Q(\varphi,w)=\((\log w)^2 + \varphi^2\)^4 \, .
\ee
Notice that the exponential factor in the integral displays a very large and sharp maximum in $\varphi=0$ in the $w\to1^-$ limit, with
\be
\left . \frac{\partial P(\varphi, w)}{\partial \varphi} \right |_{\varphi = 0} = 0 \, , \quad \left . \frac{\partial^2 P(\varphi, w)}{\partial \varphi^2} \right |_{\varphi = 0} = - \frac{8\pi^2}{(-\log w)^3} < 0 \, , \quad w < 1 \, .
\ee
This is consistent with the asymptotic expansion reported above in \eqref{thetasundborg}.

\begin{figure}[t]
\centering
{\includegraphics[width=0.598\textwidth]{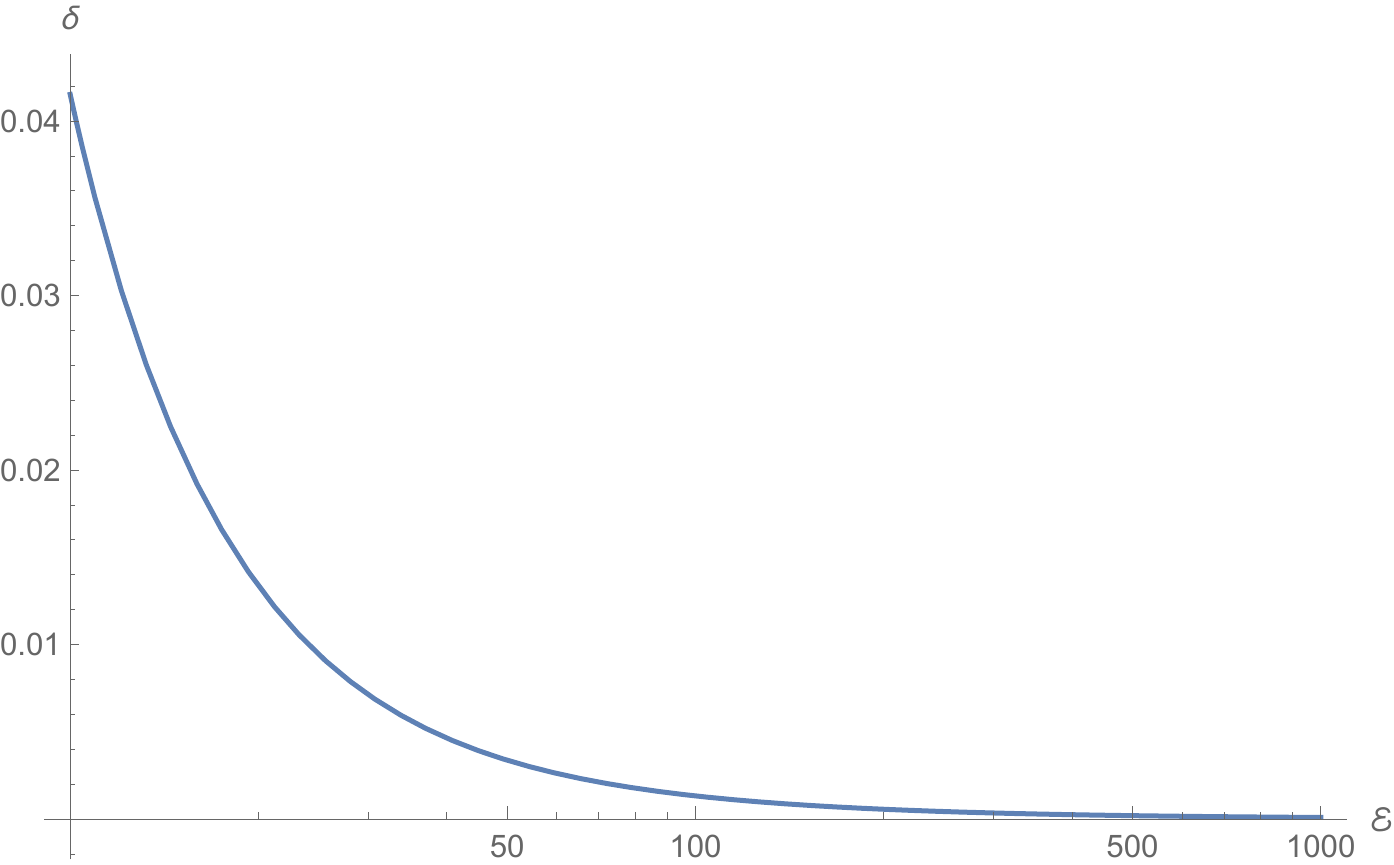}}
\caption{Plot of the relative error $\delta=\delta(w(\mathcal{E}))$ defined in \eqref{relerr} for $w=e^{-2\pi/\sqrt{\mathcal{E}}}$ as the eigenvalue $\mathcal{E}$ of the light-cone Hamiltonian varies.} 
\label{relerrplot}
\end{figure}
We can thus estimate the integral through the Laplace method, which gives
\be
\frac{1}{2\pi} \int_{-\pi}^{+\pi} \hspace{-4pt} d\varphi \,\, \Pi(\varphi,w) \sim \frac{1}{2^{18}\pi^9 \sqrt{\pi}} (-\log w)^{19/2} \, e^{-4\pi^2/\log w} \, , \quad w\to1^- \, .
\ee
Given the asymptotic expressions in  \eqref{asymtheta}, we conclude that formula \eqref{ansatz} is verified.
Moreover, we can also provide a numerical evidence. Let us define the relative error
\be
\label{relerr}
\delta(w) = \frac{\log\[\frac{1}{2\pi} \int_{0}^{2\pi} \hspace{-4pt} d\varphi \,\, \Pi(\varphi,w)\]- \log\[\frac14 \theta_4^{-16}(0,w) \, \theta_2^{-3}(0,w)\]}{\log\[\frac{1}{2\pi} \int_{0}^{2\pi} \hspace{-4pt} d\varphi \,\, \Pi(\varphi,w)\]} \, .
\ee
Its numerical analysis is shown in figure \ref{relerrplot}: $\delta$ gets smaller and smaller as $w$ approaches 1 from below.

Now, let us suppose that $10-D$ of the flat directions have been compactified. As we noted in section \ref{sec:torcom}, we just have to modify the generating function $\Pi$ as in \eqref{Pishift}. Crucially, the presence of $\mathcal K$ does not affect the above discussion about the location and the sharpness of the maximum of the integrand. Indeed, it is just a sum of trigonometric functions. As a consequence, we can just multiply the final result by
\be
\mathcal K (0, w) =\prod_{k=1}^{10-D} \theta_3\(0, w^{ \frac{R_k^2}{2\,\alpha'}}\) \theta_3\(0,w^{\frac{\alpha'}{2\,R_k^2}}\) \, ,
\ee
where
\be
\prod_{k=1}^{10-D} \theta_3\(0, w^{\frac{R_k^2}{2\,\alpha'}}\) \theta_3\(0,w^{\frac{\alpha'}{2\,R_k^2}}\) = \prod_{k=1}^{10-D} \sum_{w_k,m_k \in \mathds Z}  \(w^{\frac{R_k^2}{2\,\alpha'}}\)^{\hspace{-3pt}w_k^2} \(w^{\frac{\alpha'}{2\,R_k^2}}\)^{\hspace{-3pt}m_k^2} \, .
\ee
Using the asymptotic expansion in \eqref{asymtheta3}, we conclude that
\be \label{intKPi}
\frac{1}{2\pi} \int_{-\pi}^{+\pi} \hspace{-4pt} d\varphi \,\, \mathcal K \(\varphi,w\) \Pi(\varphi,w) \sim \frac{(2\pi)^{10-D}}{2^{18}\pi^9 \sqrt{\pi}} (-\log w)^{\frac{19}{2}-(10-D)} \, e^{-4\pi^2/\log w} \, , \quad w \to 1^- \, ,
\ee
reproduces formula \eqref{ansatzcompact}.

\newpage
\bibliographystyle{utphys}

\begin{thebibliography}{10}

\bibitem{Hagedorn:1965st}
R.~Hagedorn, ``{Statistical thermodynamics of strong interactions at
  high-energies},'' {\em Nuovo Cim. Suppl.} {\bfseries 3} (1965) 147--186.

\bibitem{Frautschi:1971ij}
S.~C. Frautschi, ``{Statistical bootstrap model of hadrons},''
  \href{http://dx.doi.org/10.1103/PhysRevD.3.2821}{{\em Phys. Rev. D}
  {\bfseries 3} (1971) 2821--2834}.

\bibitem{Carlitz:1972uf}
R.~D. Carlitz, ``{Hadronic matter at high density},''
  \href{http://dx.doi.org/10.1103/PhysRevD.5.3231}{{\em Phys. Rev. D}
  {\bfseries 5} (1972) 3231--3242}.

\bibitem{Huang:1970iq}
K.~Huang and S.~Weinberg, ``{Ultimate temperature and the early universe},''
  \href{http://dx.doi.org/10.1103/PhysRevLett.25.895}{{\em Phys. Rev. Lett.}
  {\bfseries 25} (1970) 895--897}.

\bibitem{Sundborg:1984uk}
B.~Sundborg, ``{Thermodynamics of Superstrings at High-energy Densities},''
  \href{http://dx.doi.org/10.1016/0550-3213(85)90235-4}{{\em Nucl. Phys. B}
  {\bfseries 254} (1985) 583--592}.

\bibitem{Bowick:1985az}
M.~J. Bowick and L.~C.~R. Wijewardhana, ``{Superstrings at High Temperature},''
  \href{http://dx.doi.org/10.1103/PhysRevLett.54.2485}{{\em Phys. Rev. Lett.}
  {\bfseries 54} (1985) 2485}.

\bibitem{Tye:1985jv}
S.~H.~H. Tye, ``{The Limiting Temperature Universe and Superstring},''
  \href{http://dx.doi.org/10.1016/0370-2693(85)90438-1}{{\em Phys. Lett. B}
  {\bfseries 158} (1985) 388--392}.

\bibitem{Matsuo:1986es}
N.~Matsuo, ``{Superstring Thermodynamics and Its Application to Cosmology},''
  \href{http://dx.doi.org/10.1007/BF01579145}{{\em Z. Phys. C} {\bfseries 36}
  (1987) 289}.

\bibitem{Russo:2002rq}
J.~G. Russo and A.~A. Tseytlin, ``{On solvable models of type 2B superstring in
  NS NS and RR plane wave backgrounds},''
  \href{http://dx.doi.org/10.1088/1126-6708/2002/04/021}{{\em JHEP} {\bfseries
  04} (2002) 021}, \href{http://arxiv.org/abs/hep-th/0202179}{{\ttfamily
  arXiv:hep-th/0202179}}.

\bibitem{Tseytlin:1995fh}
A.~A. Tseytlin, ``{Exact solutions of closed string theory},''
  \href{http://dx.doi.org/10.1088/0264-9381/12/10/003}{{\em Class. Quant.
  Grav.} {\bfseries 12} (1995) 2365--2410},
  \href{http://arxiv.org/abs/hep-th/9505052}{{\ttfamily arXiv:hep-th/9505052}}.

\bibitem{Metsaev:2001bj}
R.~R. Metsaev, ``{Type IIB Green-Schwarz superstring in plane wave
  Ramond-Ramond background},''
  \href{http://dx.doi.org/10.1016/S0550-3213(02)00003-2}{{\em Nucl. Phys. B}
  {\bfseries 625} (2002) 70--96},
  \href{http://arxiv.org/abs/hep-th/0112044}{{\ttfamily arXiv:hep-th/0112044}}.

\bibitem{Metsaev:2002re}
R.~R. Metsaev and A.~A. Tseytlin, ``{Exactly solvable model of superstring in
  Ramond-Ramond plane wave background},''
  \href{http://dx.doi.org/10.1103/PhysRevD.65.126004}{{\em Phys. Rev. D}
  {\bfseries 65} (2002) 126004},
  \href{http://arxiv.org/abs/hep-th/0202109}{{\ttfamily arXiv:hep-th/0202109}}.

\bibitem{Blau:2002mw}
M.~Blau, J.~M. Figueroa-O'Farrill, and G.~Papadopoulos, ``{Penrose limits,
  supergravity and brane dynamics},''
  \href{http://dx.doi.org/10.1088/0264-9381/19/18/310}{{\em Class. Quant.
  Grav.} {\bfseries 19} (2002) 4753},
  \href{http://arxiv.org/abs/hep-th/0202111}{{\ttfamily arXiv:hep-th/0202111}}.

\bibitem{Horowitz:1989bv}
G.~T. Horowitz and A.~R. Steif, ``{Space-Time Singularities in String
  Theory},'' \href{http://dx.doi.org/10.1103/PhysRevLett.64.260}{{\em Phys.
  Rev. Lett.} {\bfseries 64} (1990) 260}.

\bibitem{Horowitz:1990sr}
G.~T. Horowitz and A.~R. Steif, ``{Strings in Strong Gravitational Fields},''
  \href{http://dx.doi.org/10.1103/PhysRevD.42.1950}{{\em Phys. Rev. D}
  {\bfseries 42} (1990) 1950--1959}.

\bibitem{Ehlers:1962zz}
J.~Ehlers and W.~Kundt, ``{Exact solutions of the gravitational field
  equations},''.

\bibitem{Stephani:2003tm}
H.~Stephani, D.~Kramer, M.~A.~H. MacCallum, C.~Hoenselaers, and E.~Herlt,
  \href{http://dx.doi.org/10.1017/CBO9780511535185}{{\em {Exact solutions of
  Einstein's field equations}}}.
\newblock Cambridge Monographs on Mathematical Physics. Cambridge Univ. Press,
  Cambridge, 2003.

\bibitem{Bicak:2000ea}
J.~Bicak, ``{Selected solutions of Einstein's field equations: Their role in
  general relativity and astrophysics},'' {\em Lect. Notes Phys.} {\bfseries
  540} (2000) 1--126, \href{http://arxiv.org/abs/gr-qc/0004016}{{\ttfamily
  arXiv:gr-qc/0004016}}.

\bibitem{Amati:1988ww}
D.~Amati and C.~Klimcik, ``{Strings in a Shock Wave Background and Generation
  of Curved Geometry from Flat Space String Theory},''
  \href{http://dx.doi.org/10.1016/0370-2693(88)90355-3}{{\em Phys. Lett. B}
  {\bfseries 210} (1988) 92--96}.

\bibitem{Amati:1988sa}
D.~Amati and C.~Klimcik, ``{Nonperturbative Computation of the Weyl Anomaly for
  a Class of Nontrivial Backgrounds},''
  \href{http://dx.doi.org/10.1016/0370-2693(89)91092-7}{{\em Phys. Lett. B}
  {\bfseries 219} (1989) 443--447}.

\bibitem{Penrose1976}
R.~Penrose, {\em Any Space-Time has a Plane Wave as a Limit},
  \href{http://dx.doi.org/10.1007/978-94-010-1508-0_23}{pp.~271--275}.
\newblock Springer Netherlands, Dordrecht, 1976.
\newblock \url{https://doi.org/10.1007/978-94-010-1508-0_23}.

\bibitem{Kowalski-Glikman:1984qtj}
J.~Kowalski-Glikman, ``{Vacuum States in Supersymmetric Kaluza-Klein Theory},''
  \href{http://dx.doi.org/10.1016/0370-2693(84)90669-5}{{\em Phys. Lett. B}
  {\bfseries 134} (1984) 194--196}.

\bibitem{Gueven:2000ru}
R.~Gueven, ``{Plane wave limits and T duality},''
  \href{http://dx.doi.org/10.1016/S0370-2693(00)00517-7}{{\em Phys. Lett. B}
  {\bfseries 482} (2000) 255--263},
  \href{http://arxiv.org/abs/hep-th/0005061}{{\ttfamily arXiv:hep-th/0005061}}.

\bibitem{Blau:2002dy}
M.~Blau, J.~M. Figueroa-O'Farrill, C.~Hull, and G.~Papadopoulos, ``{Penrose
  limits and maximal supersymmetry},''
  \href{http://dx.doi.org/10.1088/0264-9381/19/10/101}{{\em Class. Quant.
  Grav.} {\bfseries 19} (2002) L87--L95},
  \href{http://arxiv.org/abs/hep-th/0201081}{{\ttfamily arXiv:hep-th/0201081}}.

\bibitem{Figueroa-OFarrill:2001hal}
J.~M. Figueroa-O'Farrill and G.~Papadopoulos, ``{Homogeneous fluxes, branes and
  a maximally supersymmetric solution of M theory},''
  \href{http://dx.doi.org/10.1088/1126-6708/2001/08/036}{{\em JHEP} {\bfseries
  08} (2001) 036}, \href{http://arxiv.org/abs/hep-th/0105308}{{\ttfamily
  arXiv:hep-th/0105308}}.

\bibitem{Blau:2001ne}
M.~Blau, J.~M. Figueroa-O'Farrill, C.~Hull, and G.~Papadopoulos, ``{A New
  maximally supersymmetric background of IIB superstring theory},''
  \href{http://dx.doi.org/10.1088/1126-6708/2002/01/047}{{\em JHEP} {\bfseries
  01} (2002) 047}, \href{http://arxiv.org/abs/hep-th/0110242}{{\ttfamily
  arXiv:hep-th/0110242}}.

\bibitem{Berenstein:2002jq}
D.~E. Berenstein, J.~M. Maldacena, and H.~S. Nastase, ``{Strings in flat space
  and pp waves from N=4 superYang-Mills},''
  \href{http://dx.doi.org/10.1088/1126-6708/2002/04/013}{{\em JHEP} {\bfseries
  04} (2002) 013}, \href{http://arxiv.org/abs/hep-th/0202021}{{\ttfamily
  arXiv:hep-th/0202021}}.

\bibitem{Mitchell:1987th}
D.~Mitchell and N.~Turok, ``{Statistical Properties of Cosmic Strings},''
  \href{http://dx.doi.org/10.1016/0550-3213(87)90626-2}{{\em Nucl. Phys. B}
  {\bfseries 294} (1987) 1138--1163}.

\bibitem{Polchinski:1985zf}
J.~Polchinski, ``{Evaluation of the One Loop String Path Integral},''
  \href{http://dx.doi.org/10.1007/BF01210791}{{\em Commun. Math. Phys.}
  {\bfseries 104} (1986) 37}.

\bibitem{OBrien:1987kzw}
K.~H. O'Brien and C.~I. Tan, ``{Modular Invariance of Thermopartition Function
  and Global Phase Structure of Heterotic String},''
  \href{http://dx.doi.org/10.1103/PhysRevD.36.1184}{{\em Phys. Rev. D}
  {\bfseries 36} (1987) 1184}.

\bibitem{McClain:1986id}
B.~McClain and B.~D.~B. Roth, ``{Modular Invariance for Interacting Bosonic
  Strings at Finite Temperature},''
  \href{http://dx.doi.org/10.1007/BF01219073}{{\em Commun. Math. Phys.}
  {\bfseries 111} (1987) 539}.

\bibitem{Alvarez:1985fw}
E.~Alvarez, ``{STRINGS AT FINITE TEMPERATURE},''
  \href{http://dx.doi.org/10.1016/0550-3213(86)90514-6}{{\em Nucl. Phys. B}
  {\bfseries 269} (1986) 596--620}. [Erratum: Nucl.Phys.B 279, 828--829
  (1987)].

\bibitem{Alvarez:1986sj}
E.~Alvarez and M.~A.~R. Osorio, ``{Superstrings at Finite Temperature},''
  \href{http://dx.doi.org/10.1103/PhysRevD.36.1175}{{\em Phys. Rev. D}
  {\bfseries 36} (1987) 1175}.

\bibitem{Atick:1988si}
J.~J. Atick and E.~Witten, ``{The Hagedorn Transition and the Number of Degrees
  of Freedom of String Theory},''
  \href{http://dx.doi.org/10.1016/0550-3213(88)90151-4}{{\em Nucl. Phys. B}
  {\bfseries 310} (1988) 291--334}.

\bibitem{Kogan:1987jd}
Y.~I. Kogan, ``{Vortices on the World Sheet and String's Critical Dynamics},''
  {\em JETP Lett.} {\bfseries 45} (1987) 709--712.

\bibitem{Sathiapalan:1986db}
B.~Sathiapalan, ``{Vortices on the String World Sheet and Constraints on Toral
  Compactification},'' \href{http://dx.doi.org/10.1103/PhysRevD.35.3277}{{\em
  Phys. Rev. D} {\bfseries 35} (1987) 3277}.

\bibitem{Deo:1989bv}
N.~Deo, S.~Jain, and C.-I. Tan, ``{STRING STATISTICAL MECHANICS ABOVE HAGEDORN
  ENERGY DENSITY},'' \href{http://dx.doi.org/10.1103/PhysRevD.40.2626}{{\em
  Phys. Rev. D} {\bfseries 40} (1989) 2626}.

\bibitem{Deo:1988jj}
N.~Deo, S.~Jain, and C.-I. Tan, ``{Strings at High-energy Densities and Complex
  Temperature},'' \href{http://dx.doi.org/10.1016/0370-2693(89)90024-5}{{\em
  Phys. Lett. B} {\bfseries 220} (1989) 125--132}.

\bibitem{Bowick:1989us}
M.~J. Bowick and S.~B. Giddings, ``{HIGH TEMPERATURE STRINGS},''
  \href{http://dx.doi.org/10.1016/0550-3213(89)90500-2}{{\em Nucl. Phys. B}
  {\bfseries 325} (1989) 631--646}.

\bibitem{Brandenberger:1988aj}
R.~H. Brandenberger and C.~Vafa, ``{Superstrings in the Early Universe},''
  \href{http://dx.doi.org/10.1016/0550-3213(89)90037-0}{{\em Nucl. Phys. B}
  {\bfseries 316} (1989) 391--410}.

\bibitem{Mertens:2015ola}
T.~G. Mertens, {\em {Hagedorn String Thermodynamics in Curved Spacetimes and
  near Black Hole Horizons}}.
\newblock PhD thesis, Gent U., 2015.
\newblock \href{http://arxiv.org/abs/1506.07798}{{\ttfamily arXiv:1506.07798
  [hep-th]}}.

\bibitem{PandoZayas:2002hh}
L.~A. Pando~Zayas and D.~Vaman, ``{Strings in RR plane wave background at
  finite temperature},''
  \href{http://dx.doi.org/10.1103/PhysRevD.67.106006}{{\em Phys. Rev. D}
  {\bfseries 67} (2003) 106006},
  \href{http://arxiv.org/abs/hep-th/0208066}{{\ttfamily arXiv:hep-th/0208066}}.

\bibitem{Greene:2002cd}
B.~R. Greene, K.~Schalm, and G.~Shiu, ``{On the Hagedorn behaviour of PP wave
  strings and N=4 SYM theory at finite R charge density},''
  \href{http://dx.doi.org/10.1016/S0550-3213(02)01071-4}{{\em Nucl. Phys. B}
  {\bfseries 652} (2003) 105--126},
  \href{http://arxiv.org/abs/hep-th/0208163}{{\ttfamily arXiv:hep-th/0208163}}.

\bibitem{Brower:2002zx}
R.~C. Brower, D.~A. Lowe, and C.-I. Tan, ``{Hagedorn transition for strings on
  pp waves and tori with chemical potentials},''
  \href{http://dx.doi.org/10.1016/S0550-3213(03)00003-8}{{\em Nucl. Phys. B}
  {\bfseries 652} (2003) 127--141},
  \href{http://arxiv.org/abs/hep-th/0211201}{{\ttfamily arXiv:hep-th/0211201}}.

\bibitem{Grignani:2003cs}
G.~Grignani, M.~Orselli, G.~W. Semenoff, and D.~Trancanelli, ``{The Superstring
  Hagedorn temperature in a pp wave background},''
  \href{http://dx.doi.org/10.1088/1126-6708/2003/06/006}{{\em JHEP} {\bfseries
  06} (2003) 006}, \href{http://arxiv.org/abs/hep-th/0301186}{{\ttfamily
  arXiv:hep-th/0301186}}.

\bibitem{Aharony:2003sx}
O.~Aharony, J.~Marsano, S.~Minwalla, K.~Papadodimas, and M.~Van~Raamsdonk,
  ``{The Hagedorn - deconfinement phase transition in weakly coupled large N
  gauge theories},'' \href{http://dx.doi.org/10.4310/ATMP.2004.v8.n4.a1}{{\em
  Adv. Theor. Math. Phys.} {\bfseries 8} (2004) 603--696},
  \href{http://arxiv.org/abs/hep-th/0310285}{{\ttfamily arXiv:hep-th/0310285}}.

\bibitem{Barbon:2004dd}
J.~L.~F. Barbon and E.~Rabinovici,
  \href{http://dx.doi.org/10.1142/9789812775344_0048}{``{Touring the Hagedorn
  ridge},''} in {\em {From Fields to Strings: Circumnavigating Theoretical
  Physics: A Conference in Tribute to Ian Kogan}}, pp.~1973--2008.
\newblock 8, 2004.
\newblock \href{http://arxiv.org/abs/hep-th/0407236}{{\ttfamily
  arXiv:hep-th/0407236}}.

\bibitem{Dienes:1998hx}
K.~R. Dienes, E.~Dudas, T.~Gherghetta, and A.~Riotto, ``{Cosmological phase
  transitions and radius stabilization in higher dimensions},''
  \href{http://dx.doi.org/10.1016/S0550-3213(98)00855-4}{{\em Nucl. Phys. B}
  {\bfseries 543} (1999) 387--422},
  \href{http://arxiv.org/abs/hep-ph/9809406}{{\ttfamily arXiv:hep-ph/9809406}}.

\bibitem{Grignani:1999sp}
G.~Grignani and G.~W. Semenoff, ``{Thermodynamic partition function of matrix
  superstrings},'' \href{http://dx.doi.org/10.1016/S0550-3213(99)00519-2}{{\em
  Nucl. Phys. B} {\bfseries 561} (1999) 243--272},
  \href{http://arxiv.org/abs/hep-th/9903246}{{\ttfamily arXiv:hep-th/9903246}}.

\bibitem{Mitchell:1987hr}
D.~Mitchell and N.~Turok, ``{Statistical Mechanics of Cosmic Strings},''
  \href{http://dx.doi.org/10.1103/PhysRevLett.58.1577}{{\em Phys. Rev. Lett.}
  {\bfseries 58} (1987) 1577}.

\bibitem{Salomonson:1985eq}
P.~Salomonson and B.-S. Skagerstam, ``{ON SUPERDENSE SUPERSTRING GASES: A
  HERETIC STRING MODEL APPROACH},''
  \href{http://dx.doi.org/10.1016/0550-3213(86)90158-6}{{\em Nucl. Phys. B}
  {\bfseries 268} (1986) 349--361}.

\bibitem{Lowe:1994nm}
D.~A. Lowe and L.~Thorlacius, ``{Hot string soup},''
  \href{http://dx.doi.org/10.1103/PhysRevD.51.665}{{\em Phys. Rev. D}
  {\bfseries 51} (1995) 665--670},
  \href{http://arxiv.org/abs/hep-th/9408134}{{\ttfamily arXiv:hep-th/9408134}}.

\bibitem{Abel:1999rq}
S.~A. Abel, J.~L.~F. Barbon, I.~I. Kogan, and E.~Rabinovici, ``{String
  thermodynamics in D-brane backgrounds},''
  \href{http://dx.doi.org/10.1088/1126-6708/1999/04/015}{{\em JHEP} {\bfseries
  04} (1999) 015}, \href{http://arxiv.org/abs/hep-th/9902058}{{\ttfamily
  arXiv:hep-th/9902058}}.

\bibitem{Abel:1999dy}
S.~A. Abel, J.~L.~F. Barbon, I.~I. Kogan, and E.~Rabinovici,
  \href{http://dx.doi.org/10.1142/9789812793850_0031}{``{Some thermodynamical
  aspects of string theory},''} in {\em {Conference on Fundamental Interactions
  from Symmetries to Black Holes (EnglerFest)}}, pp.~611--626.
\newblock 3, 1999.
\newblock \href{http://arxiv.org/abs/hep-th/9911004}{{\ttfamily
  arXiv:hep-th/9911004}}.

\bibitem{Barbon:1998ix}
J.~L.~F. Barbon and E.~Rabinovici, ``{Extensivity versus holography in anti-de
  Sitter spaces},'' \href{http://dx.doi.org/10.1016/S0550-3213(98)00824-4}{{\em
  Nucl. Phys. B} {\bfseries 545} (1999) 371--384},
  \href{http://arxiv.org/abs/hep-th/9805143}{{\ttfamily arXiv:hep-th/9805143}}.

\bibitem{Barbon:1998cr}
J.~L.~F. Barbon, I.~I. Kogan, and E.~Rabinovici, ``{On stringy thresholds in
  SYM / AdS thermodynamics},''
  \href{http://dx.doi.org/10.1016/S0550-3213(98)00868-2}{{\em Nucl. Phys. B}
  {\bfseries 544} (1999) 104--144},
  \href{http://arxiv.org/abs/hep-th/9809033}{{\ttfamily arXiv:hep-th/9809033}}.

\bibitem{Barbon:2001di}
J.~L.~F. Barbon and E.~Rabinovici, ``{Closed string tachyons and the Hagedorn
  transition in AdS space},''
  \href{http://dx.doi.org/10.1088/1126-6708/2002/03/057}{{\em JHEP} {\bfseries
  03} (2002) 057}, \href{http://arxiv.org/abs/hep-th/0112173}{{\ttfamily
  arXiv:hep-th/0112173}}.

\bibitem{Alvarez:1984ee}
E.~Alvarez, ``{SUPERSTRING COSMOLOGY},''
  \href{http://dx.doi.org/10.1103/PhysRevD.31.418}{{\em Phys. Rev. D}
  {\bfseries 31} (1985) 418}. [Erratum: Phys.Rev.D 33, 1206 (1986)].

\bibitem{Deo:1991mp}
N.~Deo, S.~Jain, O.~Narayan, and C.-I. Tan, ``{The Effect of topology on the
  thermodynamic limit for a string gas},''
  \href{http://dx.doi.org/10.1103/PhysRevD.45.3641}{{\em Phys. Rev. D}
  {\bfseries 45} (1992) 3641--3650}.

\bibitem{Deo:1991af}
N.~Deo, S.~Jain, and C.-I. Tan, ``{The ideal gas of strings},'' in {\em
  {International Colloquium on Modern Quantum Field Theory}}.
\newblock 3, 1991.

\bibitem{Frey:2023khe}
A.~R. Frey, R.~Mahanta, A.~Maharana, F.~Muia, F.~Quevedo, and G.~Villa,
  ``{String thermodynamics in and out of equilibrium: Boltzmann equations and
  random walks},'' \href{http://dx.doi.org/10.1007/JHEP03(2024)112}{{\em JHEP}
  {\bfseries 03} (2024) 112}, \href{http://arxiv.org/abs/2310.11494}{{\ttfamily
  arXiv:2310.11494 [hep-th]}}.

\bibitem{Giddings:1989xe}
S.~B. Giddings, ``{Strings at the Hagedorn Temperature},''
  \href{http://dx.doi.org/10.1016/0370-2693(89)90288-8}{{\em Phys. Lett. B}
  {\bfseries 226} (1989) 55--61}.

\bibitem{Jain:1997ga}
S.~Jain, ``{Absence of initial singularities in superstring cosmology},''
  \href{http://dx.doi.org/10.1007/BF02709328}{{\em J. Astrophys. Astron.}
  {\bfseries 18} (1997) 363},
  \href{http://arxiv.org/abs/gr-qc/9708018}{{\ttfamily arXiv:gr-qc/9708018}}.

\bibitem{Drukker:2000ep}
N.~Drukker, D.~J. Gross, and A.~A. Tseytlin, ``{Green-Schwarz string in AdS(5)
  x S**5: Semiclassical partition function},''
  \href{http://dx.doi.org/10.1088/1126-6708/2000/04/021}{{\em JHEP} {\bfseries
  04} (2000) 021}, \href{http://arxiv.org/abs/hep-th/0001204}{{\ttfamily
  arXiv:hep-th/0001204}}.

\bibitem{Gautason:2021vfc}
F.~F. Gautason and V.~G.~M. Puletti, ``{Precision holography for 5D Super
  Yang-Mills},'' \href{http://dx.doi.org/10.1007/JHEP03(2022)018}{{\em JHEP}
  {\bfseries 03} (2022) 018}, \href{http://arxiv.org/abs/2111.15493}{{\ttfamily
  arXiv:2111.15493 [hep-th]}}.

\bibitem{Green:2012oqa}
M.~B. Green, J.~H. Schwarz, and E.~Witten,
  \href{http://dx.doi.org/10.1017/CBO9781139248563}{{\em {Superstring Theory
  Vol. 1}: {25th Anniversary Edition}}}.
\newblock Cambridge Monographs on Mathematical Physics. Cambridge University
  Press, 11, 2012.

\bibitem{Hyun:2003ks}
S.-j. Hyun, J.-D. Park, and S.-H. Yi, ``{Thermodynamic behavior of IIA string
  theory on a pp wave},''
  \href{http://dx.doi.org/10.1088/1126-6708/2003/11/006}{{\em JHEP} {\bfseries
  11} (2003) 006}, \href{http://arxiv.org/abs/hep-th/0304239}{{\ttfamily
  arXiv:hep-th/0304239}}.

\bibitem{Bigazzi:2003jk}
F.~Bigazzi and A.~L. Cotrone, ``{On zero point energy, stability and Hagedorn
  behavior of type IIB strings on pp waves},''
  \href{http://dx.doi.org/10.1088/1126-6708/2003/08/052}{{\em JHEP} {\bfseries
  08} (2003) 052}, \href{http://arxiv.org/abs/hep-th/0306102}{{\ttfamily
  arXiv:hep-th/0306102}}.

\bibitem{Sugawara:2002rs}
Y.~Sugawara, ``{Thermal amplitudes in DLCQ superstrings on PP waves},''
  \href{http://dx.doi.org/10.1016/S0550-3213(02)01030-1}{{\em Nucl. Phys. B}
  {\bfseries 650} (2003) 75--113},
  \href{http://arxiv.org/abs/hep-th/0209145}{{\ttfamily arXiv:hep-th/0209145}}.

\bibitem{Bigazzi:2024biz}
F.~Bigazzi, T.~Canneti, F.~Castellani, A.~L. Cotrone, and W.~M\"uck,
  ``{Hagedorn temperature in holography: world-sheet and effective
  approaches},'' \href{http://arxiv.org/abs/2407.00375}{{\ttfamily
  arXiv:2407.00375 [hep-th]}}.

\bibitem{Harmark:2024ioq}
T.~Harmark, ``{Hagedorn temperature from the thermal scalar in AdS and pp-wave
  backgrounds},'' \href{http://dx.doi.org/10.1007/JHEP06(2024)140}{{\em JHEP}
  {\bfseries 06} (2024) 140}, \href{http://arxiv.org/abs/2402.06001}{{\ttfamily
  arXiv:2402.06001 [hep-th]}}.

\bibitem{Berkooz:2007fe}
M.~Berkooz, Z.~Komargodski, and D.~Reichmann, ``{Thermal AdS(3), BTZ and
  competing winding modes condensation},''
  \href{http://dx.doi.org/10.1088/1126-6708/2007/12/020}{{\em JHEP} {\bfseries
  12} (2007) 020}, \href{http://arxiv.org/abs/0706.0610}{{\ttfamily
  arXiv:0706.0610 [hep-th]}}.

\bibitem{Hyun:2002wu}
S.-j. Hyun and H.-j. Shin, ``{N=(4,4) type 2A string theory on PP wave
  background},'' \href{http://dx.doi.org/10.1088/1126-6708/2002/10/070}{{\em
  JHEP} {\bfseries 10} (2002) 070},
  \href{http://arxiv.org/abs/hep-th/0208074}{{\ttfamily arXiv:hep-th/0208074}}.

\bibitem{Shin:2003ae}
H.-j. Shin, K.~Sugiyama, and K.~Yoshida, ``{Partition function and open /
  closed string duality in type IIA string theory on a PP wave},''
  \href{http://dx.doi.org/10.1016/j.nuclphysb.2003.07.015}{{\em Nucl. Phys. B}
  {\bfseries 669} (2003) 78--102},
  \href{http://arxiv.org/abs/hep-th/0306087}{{\ttfamily arXiv:hep-th/0306087}}.

\bibitem{Witten:1998zw}
E.~Witten, ``{Anti-de Sitter space, thermal phase transition, and confinement
  in gauge theories},''
  \href{http://dx.doi.org/10.4310/ATMP.1998.v2.n3.a3}{{\em Adv. Theor. Math.
  Phys.} {\bfseries 2} (1998) 505--532},
  \href{http://arxiv.org/abs/hep-th/9803131}{{\ttfamily arXiv:hep-th/9803131}}.

\bibitem{Bigazzi:2004ze}
F.~Bigazzi, A.~L. Cotrone, L.~Martucci, and L.~A. Pando~Zayas, ``{Wilson loop,
  Regge trajectory and hadron masses in a Yang-Mills theory from semiclassical
  strings},'' \href{http://dx.doi.org/10.1103/PhysRevD.71.066002}{{\em Phys.
  Rev. D} {\bfseries 71} (2005) 066002},
  \href{http://arxiv.org/abs/hep-th/0409205}{{\ttfamily arXiv:hep-th/0409205}}.

\bibitem{Sidorov1985LecturesOT}
Y.~V. Sidorov, M.~V. Fedoryuk, and M.~I. Shabunin, ``Lectures on the theory of
  functions of a complex variable,''
\newblock Mir Publishers, 1985.

\bibitem{Horowitz:1997jc}
G.~T. Horowitz and J.~Polchinski, ``{Selfgravitating fundamental strings},''
  \href{http://dx.doi.org/10.1103/PhysRevD.57.2557}{{\em Phys. Rev. D}
  {\bfseries 57} (1998) 2557--2563},
  \href{http://arxiv.org/abs/hep-th/9707170}{{\ttfamily arXiv:hep-th/9707170}}.

\bibitem{Bigazzi:2023oqm}
F.~Bigazzi, T.~Canneti, and W.~M\"uck, ``{Semiclassical quantization of the
  superstring and Hagedorn temperature},''
  \href{http://dx.doi.org/10.1007/JHEP08(2023)185}{{\em JHEP} {\bfseries 08}
  (2023) 185}, \href{http://arxiv.org/abs/2306.00588}{{\ttfamily
  arXiv:2306.00588 [hep-th]}}.

\bibitem{Bigazzi:2023hxt}
F.~Bigazzi, T.~Canneti, and A.~L. Cotrone, ``{Higher order corrections to the
  Hagedorn temperature at strong coupling},''
  \href{http://dx.doi.org/10.1007/JHEP10(2023)056}{{\em JHEP} {\bfseries 10}
  (2023) 056}, \href{http://arxiv.org/abs/2306.17126}{{\ttfamily
  arXiv:2306.17126 [hep-th]}}.

\bibitem{Sundborg:1999ue}
B.~Sundborg, ``{The Hagedorn transition, deconfinement and N=4 SYM theory},''
  \href{http://dx.doi.org/10.1016/S0550-3213(00)00044-4}{{\em Nucl. Phys. B}
  {\bfseries 573} (2000) 349--363},
  \href{http://arxiv.org/abs/hep-th/9908001}{{\ttfamily arXiv:hep-th/9908001}}.

\bibitem{tHooft:1984kcu}
G.~'t~Hooft, ``{On the Quantum Structure of a Black Hole},''
  \href{http://dx.doi.org/10.1016/0550-3213(85)90418-3}{{\em Nucl. Phys. B}
  {\bfseries 256} (1985) 727--745}.

\bibitem{Banks:2003vp}
T.~Banks, ``{A Critique of pure string theory: Heterodox opinions of diverse
  dimensions},'' \href{http://arxiv.org/abs/hep-th/0306074}{{\ttfamily
  arXiv:hep-th/0306074}}.

\bibitem{NIST:DLMF}
``{\it NIST Digital Library of Mathematical Functions}.''
  \url{https://dlmf.nist.gov/}, release 1.2.1 of 2024-06-15.
\newblock \url{https://dlmf.nist.gov/}. F.~W.~J. Olver, A.~B. {Olde Daalhuis},
  D.~W. Lozier, B.~I. Schneider, R.~F. Boisvert, C.~W. Clark, B.~R. Miller,
  B.~V. Saunders, H.~S. Cohl, and M.~A. McClain, eds.

\bibitem{Wolfram}
``{\it The Mathematical Functions Site}.''
  \url{https://functions.wolfram.com/}.
\newblock \url{https://functions.wolfram.com/}. O. Marichev, M. Trott, and S.
  Wolfram.

\end{thebibliography}
\providecommand{\href}[2]{#2}\begingroup\raggedright\endgroup

\end{document}